\documentclass[reprint,superscriptaddress,amsmath,amssymb,aps,longbibliography]{revtex4-2}

\usepackage{graphicx}
\usepackage{booktabs}
\usepackage{multirow}
\usepackage{hyperref}
\usepackage{lmodern}

\newcommand{\ocg}{\,\widetilde{\otimes}\,}
\DeclareMathOperator*{\argmin}{arg\,min}

\begin{document}

\title{Unconstrained learning of equivariant atomistic properties}

\author{Joseph W. Abbott}
\affiliation{Laboratory of Computational Science and Modeling, Institut des Mat\'eriaux, \'Ecole Polytechnique F\'ed\'erale de Lausanne, 1015 Lausanne, Switzerland}

\author{Paolo Pegolo}
\affiliation{Laboratory of Computational Science and Modeling, Institut des Mat\'eriaux, \'Ecole Polytechnique F\'ed\'erale de Lausanne, 1015 Lausanne, Switzerland}

\author{Pol Febrer}
\affiliation{Laboratory of Computational Science and Modeling, Institut des Mat\'eriaux, \'Ecole Polytechnique F\'ed\'erale de Lausanne, 1015 Lausanne, Switzerland}

\author{Michelangelo Domina}
\affiliation{Laboratory of Computational Science and Modeling, Institut des Mat\'eriaux, \'Ecole Polytechnique F\'ed\'erale de Lausanne, 1015 Lausanne, Switzerland}

\author{Sandra Saade}
\affiliation{Laboratory of Computational Science and Modeling, Institut des Mat\'eriaux, \'Ecole Polytechnique F\'ed\'erale de Lausanne, 1015 Lausanne, Switzerland}

\author{Matthias Kellner}
\affiliation{Laboratory of Computational Science and Modeling, Institut des Mat\'eriaux, \'Ecole Polytechnique F\'ed\'erale de Lausanne, 1015 Lausanne, Switzerland}

\author{Filippo Bigi}
\affiliation{Laboratory of Computational Science and Modeling, Institut des Mat\'eriaux, \'Ecole Polytechnique F\'ed\'erale de Lausanne, 1015 Lausanne, Switzerland}

\author{Philip Loche}
\affiliation{Laboratory of Computational Science and Modeling, Institut des Mat\'eriaux, \'Ecole Polytechnique F\'ed\'erale de Lausanne, 1015 Lausanne, Switzerland}
\affiliation{Department of Physics, Technical University of Munich}
\affiliation{Atomistic Modelling Center, Munich Data Science Institute, Technical University of Munich}
\affiliation{Munich Center for Machine Learning (MCML)}

\author{Michele Ceriotti}
\email{michele.ceriotti@epfl.ch}
\affiliation{Laboratory of Computational Science and Modeling, Institut des Mat\'eriaux, \'Ecole Polytechnique F\'ed\'erale de Lausanne, 1015 Lausanne, Switzerland}

\date{\today}

\begin{abstract}
Machine-learned interatomic potentials based on graph neural networks (GNNs) have made first-principles accuracy routine for the potential energy surface (PES). Their backbones, often O(3)-equivariant by design, are increasingly used to learn other atomistic properties.
Many of these are of high angular order or pseudotensorial character, defined per atom, or expanded on large atom-centered basis sets. Such targets strain equivariant GNNs, whose tensor products grow steeply in cost with angular order and can fail to represent the target if truncated, forcing difficult trade-offs between hyperparameters.
Unconstrained models like the Point-Edge Transformer (PET) instead replace explicit O(3) algebra with edge-to-edge attention and learn equivariance from data, so that the same backbone can in principle learn any equivariant property. We demonstrate how to use PET to learn molecular dipoles and polarizabilities, NMR chemical shielding tensors, and auxiliary-basis expansions of the electron density. Increasing the backbone size systematically improves both accuracy and equivariance error, and targeted architectural and training choices reduce the symmetry breaking further. PET matches or improves on the released reference models at lower inference cost and, compared with equivariant models, a smaller memory footprint.
The predictions give accurate IR and Raman spectra and isotropic shieldings; the predicted densities accelerate self-consistent field calculations and give one-shot ground-state properties, and pretrained density models transfer to higher levels of theory by fine-tuning on little data. The impact of approximate equivariance is small for most observables, further validating this approach to train fast, accurate, and useful surrogate models for properties beyond the PES.
\end{abstract}

\maketitle
\section{Introduction}

\begin{figure*}[t]
    \centering
    \includegraphics[width=\linewidth]{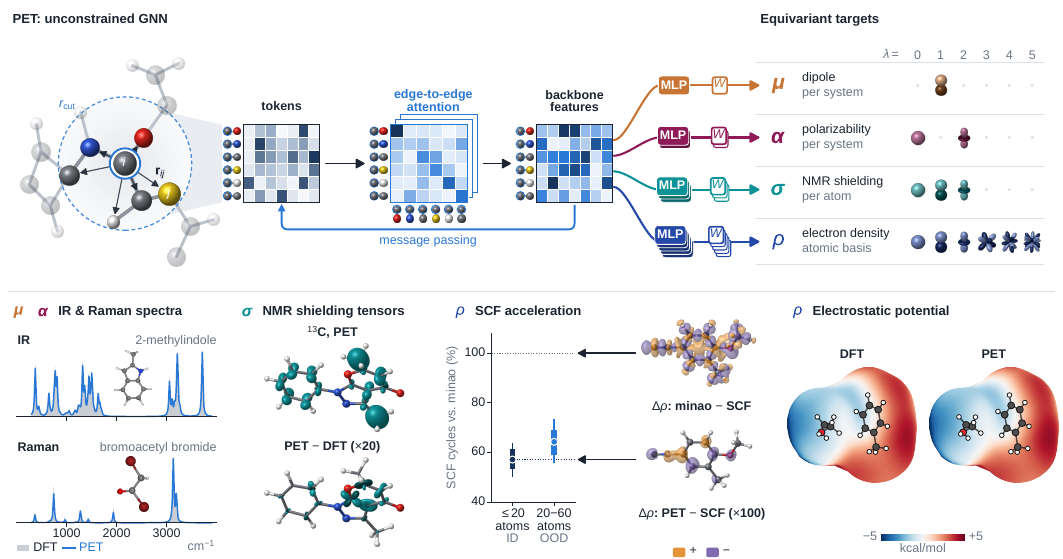}
    \caption{
    Overview of this work. Top: the unconstrained PET architecture, built on a backbone of edge-to-edge attention and message passing, with a head and readout for each O(3)-irrep block of the equivariant targets considered. Bottom: the same backbone architecture, trained on different datasets, can learn general equivariant atomistic properties, and be used to derive IR/Raman spectra, NMR chemical shielding tensors and electrostatic potentials, and to accelerate SCF calculations.
    }
    \label{fig:overview}
\end{figure*}

Machine-learned interatomic potentials (MLIPs), trained to reproduce first-principles energies, forces, and stresses, have made stable molecular dynamics at density-functional theory (DFT) accuracy a routine tool for probing the structure, thermodynamics, and dynamics of molecules and materials~\cite{behler2007neural,bartok2010gap,deringer2019mlip,batatia2022mace,rupp2012qm7,smith2017ani1,wang2019coarse,seute2025grappa,mazitov2025petmad,behler2026roadmap}.
Many of the experimentally relevant quantities, however, depend on more than the potential energy surface (PES)~\cite{ceriotti2022beyond,kulik2022roadmap}. Surrogate models for other quantum-mechanical properties likewise bypass the self-consistent field (SCF) procedure~\cite{grisafi2018sagpr,grisafi2019transfer}: e.g., molecular dipoles and polarizabilities give access to infrared (IR) and Raman spectra~\cite{gastegger2017ir,veit2020muml,raimbault2019raman,wilkins2019polarizability,schutt2021painn,gonnheimer2026macemdp,xu2025tace,martin2026macefield}, NMR chemical shielding tensors to the determination of molecular crystal structures~\cite{paruzzo2018shiftml,kellner2025shiftml3,bornes2026nmr,benmahmoud2026ssnmr}, and a predicted electron density to one-shot ground-state properties or accelerated SCF cycles~\cite{brockherde2017bypass,grisafi2023electronic,nigam2022ham,liu2026scfbench}.

Unlike the energy, which is invariant, these targets are tensorial and must transform in a prescribed way that is equivariant under rotations and inversions of the structure. They may also be of high angular order, carry pseudotensorial components, or be expanded on large atom-centered basis sets.
A model of such a target \emph{must} respect this symmetry to be physical, but how the symmetry is best respected---whether enforced on an architectural level or learned through data---remains an open question, trading exactness against cost and expressivity in ways that are still actively debated~\cite{duval2023hitchhiker,puny2022frame,langer2024broken,qu2024scalable,domina2026unconstrained,bigi2026pushing}.

The architectures used to learn properties beyond the PES have largely followed those of MLIPs, and in particular equivariant graph neural networks (GNNs)~\cite{batzner2022nequip,batatia2022mace}, which build this symmetry in exactly through Clebsch-Gordan (CG) tensor products.
These operations have a cost that scales steeply with angular order~\cite{batatia2025design}. In practice, the features are truncated at an angular order possibly below that of the target, which is then reached through products of lower-order features~\cite{qian2026equivariant}. Faster formulations of the tensor product reduce this cost~\cite{passaro2023escn, luo2024gaunt, zaverkin2024ictp, cheng2024cartesian, xu2025tace, xu2026tece}, but they still require a cutoff in angular order to be practical, with truncation limiting what the model can learn in some cases~\cite{cen2024highdegree}. In fact, the more complex the symmetry of the target, the more expensive it is for an exactly equivariant model to represent it.

Unconstrained GNNs such as the Point-Edge Transformer (PET) challenge this paradigm: discarding roto-equivariance as an architectural constraint altogether and replacing CG tensor products with simpler, faster, and more expressive attention between the edges of each atomic neighborhood, PET instead learns the symmetry from data through rotational augmentation~\cite{pozdnyakov2023pet}.
This choice is motivated by: \textit{i}) the growing availability of large, high-quality reference datasets~\cite{ramakrishnan2014qm9,hoja2021qm7x,eastman2023spice,ganscha2025qcml,barrosoluque2026omat24,mazitov2025mad,levine2025omol25,malosso2026datasets,chorna2026hierarchical} (far larger than those available when the data \textit{efficiency} of equivariant models in the low-data regime was a leading design criterion), and \textit{ii}) the observation that sufficiently flexible models do learn physical symmetries when trained appropriately~\cite{langer2024broken,qu2024scalable,neumann2024orb,kreiman2025graph,mazitov2025petmad}, with the benefit of physics-based inductive biases diminishing or even reversing as data and model size grow~\cite{qu2026lrattention}.
PET has already proven competitive as a universal MLIP~\cite{mazitov2025petmad,bigi2026pushing}, and has been applied to other specific tasks such as learning molecular dipoles~\cite{pozdnyakov2023pet}, chemical shielding tensors~\cite{kellner2025shiftml3,kellner2026shiftml4}, electronic density of states~\cite{how2026petmaddos}, long-stride molecular dynamics~\cite{bigi2025flashmd} and excitation energies~\cite{malosso2026excited}.
It was recently shown how symmetries develop during training, and that PET can learn the symmetries of the PES to high accuracy, exposing also where it can fail for isolated high-order or pseudotensorial targets, and how an easy fix can be applied to the former of those cases~\cite{domina2026unconstrained}. These applications each adapted PET to a single target, however, and how one backbone performs across general equivariant targets, up to high angular order and in practical scenarios, how its symmetry breaking can be controlled, and how it propagates to observables, has not been studied systematically.
Several open questions remain: Is learned, approximate symmetry good enough for properties with a more complicated structure than the PES? For such targets, can it be controlled and how? In practice, what can an unconstrained model offer over its equivariant counterparts when given the task of learning general equivariant targets?

In this work, we aim to address these points using molecular dipoles and polarizabilities, NMR chemical shielding tensors, and basis-set expansions of the electron density as targets (Fig.~\ref{fig:overview}). We demonstrate that PET, despite being unconstrained, can learn practically useful targets with complex symmetries, and may offer a better trade-off between accuracy and cost than exactly equivariant ones.
We start in Sec.~\ref{sec:symmetries} by investigating the incompleteness of the basis many equivariant GNNs are built on, and how PET's unconstrained symmetry alleviates this.
Next, in Sec.~\ref{sec:learning} we demonstrate that the same backbone can be used to learn all of the properties considered regardless of their O(3) symmetry, that scaling its size under simple heuristics systematically improves accuracy and equivariance errors, and that the remaining equivariance error can be reduced further.
Finally in Sec.~\ref{sec:validation} we validate the resulting models on IR and Raman spectra, NMR isotropic chemical shieldings, and DFT calculations initialized from predicted densities, comparing performance to released reference models in terms of speed, accuracy, and memory footprint, and contextualize their approximate equivariance within relevant downstream physical validation tasks.

\section{Learning symmetries}
\label{sec:symmetries}

\begin{figure}
    \centering
    \includegraphics[width=\linewidth]{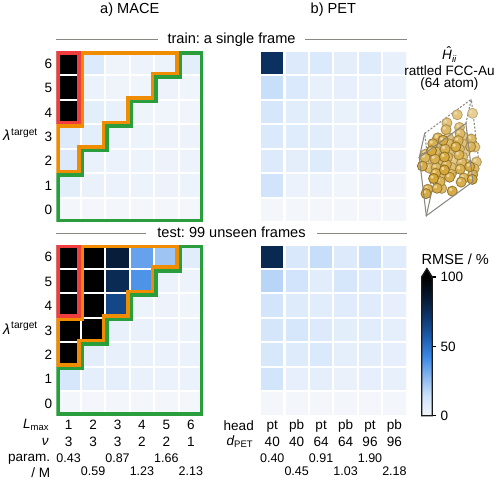}
    \caption{
    Relative RMSE, normalized by the un-centered RMS of the reference so that 100\% corresponds to predicting zero, of a) equivariant (MACE) and b) unconstrained (PET) models of matched parameter count on the on-site Hamiltonian blocks $\hat{H}_{ii}$ of rattled FCC-Au~\cite{qian2026equivariant}, by target angular order, when fitted to a single 64-atom structure (top) and tested on 99 unseen ones (bottom). Columns are labeled by the hyperparameters of each model and its number of parameters in millions: the spherical expansion order $L_{\text{max}}$ and correlation order $\nu$ for MACE, and the head type, per-target (`pt') or per-block (`pb'), and edge feature dimension $d_{\text{PET}}$ for PET. Outlines mark the MACE regimes: target represented directly, $L_{\text{max}} \ge \lambda^{\text{target}}$ (green); reached only through the product basis of Eq.~\eqref{eq:product_basis}, $L_{\text{max}} < \lambda^{\text{target}} \le L_{\text{max}} \times \nu$ (orange); beyond the reach of the product basis, $\lambda^{\text{target}} > L_{\text{max}} \times \nu$ (red). The $\lambda = 0$ block is almost entirely removed by the per-species baseline, so its relative error is close to zero for every model.
    }
    \label{fig:angular_order_benchmark}
\end{figure}

\subsection{Equivariant targets and O(3) symmetry}

Consider a tensorial target $y_{(i)}$, either defined for the whole system or per atom $i$. Applying a transformation $g \in \mathrm{O}(3)$ to the input structure---a proper rotation $R \in \mathrm{SO}(3)$, optionally composed with inversion, so that $\det g = \pm 1$---transforms the target according to a fixed linear representation set by its tensorial character. In the Cartesian basis a rank-$r$ tensor transforms as
\begin{equation}
    y_{a_1 \dots a_r} \rightarrow (\det g)^{p} \, g_{a_1 b_1} \cdots g_{a_r b_r} \, y_{b_1 \dots b_r},
    \label{eq:cartesian_transform}
\end{equation}
with $p = 0$ for proper tensors and $p = 1$ for pseudotensors. This representation is in general reducible. Through a fixed change of basis, any such tensor can be rewritten as a direct sum of \textit{irreducible representations}, or \textit{irreps}, of $\mathrm{O}(3)$, labeled by $\alpha = (\lambda, \sigma)$, where $\lambda = 0, 1, 2, \dots$ is the angular order and $\sigma = \pm 1$ distinguishes proper ($\sigma = +1$) from pseudotensorial ($\sigma = -1$) irreps:
\begin{align}
    y_{(i)} \leftrightarrow \bigoplus_{\alpha} y^{\alpha}_{(i) \mu},
    \qquad
    y^{\alpha}_{(i) \mu} &\rightarrow \sum_{\mu'} \rho^{\alpha}_{\mu \mu'}(g) \, y^{\alpha}_{(i) \mu'},
    \label{eq:irrep_transform}
\end{align}
where $\mu = -\lambda, \dots, \lambda$ indexes the $2\lambda + 1$ components of each irrep. In this spherical basis each block transforms independently of the others: under a proper rotation with the Wigner D-matrix $D^{\lambda}(R)$, and under inversion by acquiring the sign $\sigma (-1)^{\lambda}$.
Scalars such as the energy are proper invariants $(0, +1)$, the dipole and forces are proper vectors $(1, +1)$, the polarizability and stress decompose as $(0, +1) \oplus (2, +1)$, a general rank-2 tensor such as the NMR chemical shielding additionally carries an antisymmetric pseudovector $(1, -1)$ block, and an atom-centered basis-set expansion of the electron density contributes one proper irrep $(\lambda, +1)$ for each basis function of angular momentum $\lambda$. The set of irreps a target spans determines the symmetry a model must reproduce.

\begin{figure*}[t]
    \centering
    \includegraphics[width=\textwidth]{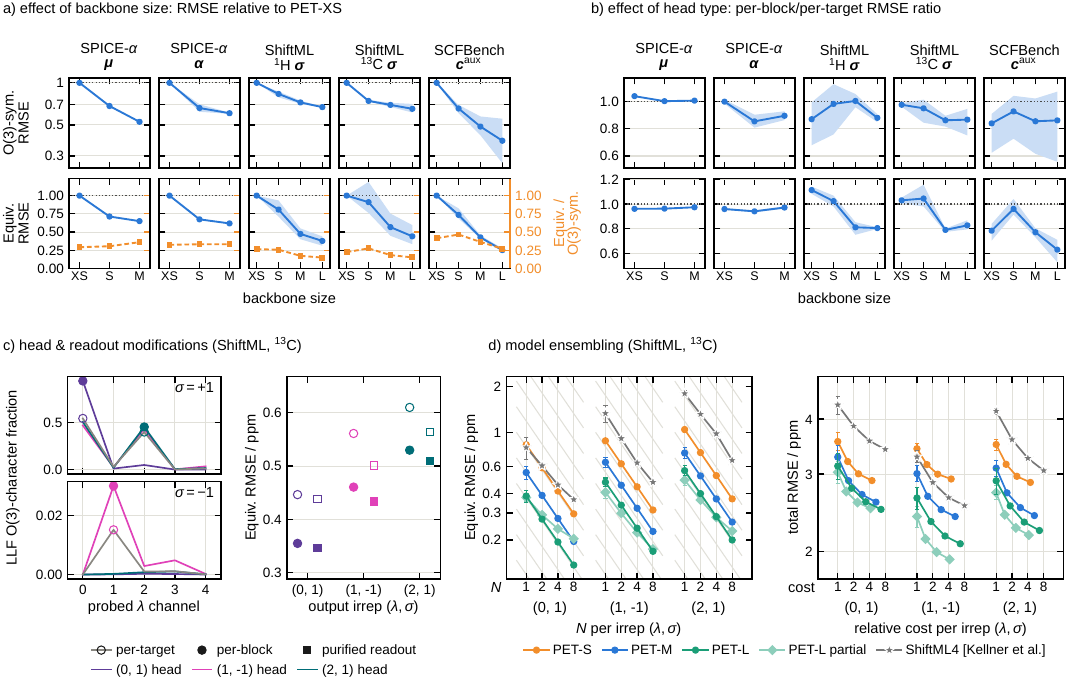}
    \caption{
    Accuracy and equivariance of PET models, in the spherical basis. a), b) $\mathrm{O}(3)$-symmetrized RMSE (top) and equivariance RMSE (bottom) of each target, as the geometric mean over its irreps (line) and their range (band): a) per-block models of increasing backbone size, relative to \texttt{PET-XS}, with the ratio of equivariance to symmetrized RMSE on the right axis (orange, dashed); b) ratio of per-block to per-target heads at each backbone size. SCFBench models use atom type-conditioned readouts. Per-irrep values are given in Fig.~\ref{fig:si_size_ablation_irreps}. c), d) $^{13}$C chemical shieldings. c) Head type, for \texttt{PET-L}: $\mathrm{O}(3)$-character fraction of the LLF of each head by probed angular order, for proper ($\sigma = +1$) and pseudotensorial ($\sigma = -1$) characters, with markers at the irreps each head targets; and equivariance RMSE for each target irrep, before and after linear purification of the readout. d) Ensembles of 1, 2, 4 and 8 members: equivariance RMSE, with $N^{-1/2}$ guide lines, and total RMSE against inference cost relative to a single model, for deep ensembles of PET and of the ShiftML4 architecture~\cite{kellner2026shiftml4}, and partial \texttt{PET-L} ensembles. The cost of a ShiftML4 ensemble is taken as $N$ times that of one member.
    }
    \label{fig:equivariance_overview}
\end{figure*}

\subsection{Equivariant architectures and basis incompleteness}

A model $f$ that maps a structure $A$ (atomic positions and types, including also lattice vectors if periodic) to a prediction $\tilde{y} = f(A)$ is $\mathrm{O}(3)$-\textit{equivariant} if
\begin{equation}
    f(g A) = \rho(g) \, f(A) \quad \forall \ g \in \mathrm{O}(3),
    \label{eq:equivariance}
\end{equation}
where $\rho(g)$ is the block-diagonal direct sum of the representation $\rho^{\alpha}(g)$ of Eq.~\eqref{eq:irrep_transform} over the irreps of the target, and invariance is the special case $\rho(g) = 1$ of a $(0, +1)$ target.
Translations and permutations of identical atoms are handled trivially by operating on interatomic displacements and summing over neighbors~\cite{musil2021physics,duval2023hitchhiker}, and we do not explicitly consider them further.

Equivariant GNNs~\cite{thomas2018tensorfield,batzner2022nequip,batatia2022mace,liao2023equiformer,musaelian2023allegro,bochkarev2024grace} are restricted to operations that generate and preserve features with $\mathrm{O}(3)$ symmetry, and so satisfy Eq.~\eqref{eq:equivariance} exactly by construction.
Any nonlinear equivariant GNN expresses a target irrep block in a form equivalent to
\begin{equation}
    y^{\lambda \sigma}_{(i) \mu} = \sum_{k} c_{k} \, B^{\lambda \sigma}_{k \mu},
    \label{eq:equivariant_basis}
\end{equation}
where the $B^{\lambda \sigma}_{k}$ are tensors transforming as the $(\lambda, \sigma)$ irrep that form a basis of the target space, and the $c_{k}$ are invariant nonlinear functions of the atomic environment.
A complete basis can be constructed in principle~\cite{domina2025scalar}, but is impractically expensive for high $\lambda$.
Instead, a spherical expansion of the neighbor density of each atom is combined through CG tensor products up to a correlation order $\nu$~\cite{drautz2019ace,drautz2020tensorial}, retaining only the features that match, or contribute to, the irreps of the target:
\begin{align}
    B^{l}_{n \mu} = \sum_{j} R_{n}(r_{ij}) \, Y^{l}_{\mu}(\hat{\textbf{r}}_{ij}) \ , \\
    B^{\lambda}_{n_1 n_2 l_1 l_2 \, \mu} = \left( B^{l_1}_{n_1} \ocg B^{l_2}_{n_2} \right)_{\lambda \mu} \ .
    \label{eq:product_basis}
\end{align}
The expansion uses radial functions $R_n$ and spherical harmonics $Y^{l}_{\mu}$ truncated at some $L_{\text{max}}$; $\ocg$ denotes the CG-coupled product, of angular order $\lambda \le l_1 + l_2$, and higher correlation orders follow by iteration~\cite{nigam2020recursive,nigam2022unified}.

Auxiliary-basis expansions of the electron density~\cite{grisafi2019transfer,fabrizio2019noncovalent,lewis2021condensed,grisafi2023electronic,rackers2023cracking,fu2024recipe,qiao2022informing,liu2026scfbench,lou2026moire}, and the electronic Hamiltonian~\cite{unke2021phisnet,li2022deeph,qian2026equivariant,nigam2022ham,suman2025design} or density matrix~\cite{febrer2025graph2mat}, can require angular components up to $\lambda_{\text{max}}^{\text{target}} = 6$ or higher. For the electron density, higher-order equivariant features have been reported to improve accuracy, particularly for systems with strong angular variation~\cite{koker2024charge3net}.
Setting $L_{\text{max}} \ge \lambda_{\text{max}}^{\text{target}}$ guarantees a complete basis for the target, but is often prohibitively expensive in time and memory: naive sparse CG products in the spherical basis scale as $\mathcal{O}(L^6)$, and even with truncation schemes in the uncoupled basis as $\mathcal{O}(L^3)$~\cite{maennel2024complete,bigi2026space}.
Backbone features in the literature therefore rarely exceed $L_{\text{max}} = 4$, with higher-order targets reached through a product basis in which lower-order features are coupled by further CG products, such that $L_{\text{max}} \times \nu \ge \lambda_{\text{max}}^{\text{target}}$~\cite{qian2026equivariant,aertebjerg2026initialization,febrer2025graph2mat}.

The product basis, however, is incomplete, meaning it cannot reproduce arbitrary equivariant functions of the atomic coordinates. This is most evident for structures whose low-order spherical expansions vanish by symmetry while higher-order ones do not~\cite{cen2024highdegree}.
The most prevalent example in atomistic ML is the octahedral environment, whose first non-vanishing spherical expansion beyond the invariant has $L = 4$, as already noted in the context of learning electronic structure~\cite{nigam2026external}. With $L_{\text{max}} < 4$, no product can recover the missing terms.
The problem is not limited to vanishing expansions, but extends to structures whose spherical expansions are restricted by symmetry to fewer than their $2L + 1$ dimensions (Sec.~\ref{sec:si_dimensionalityanalysis}).
Structures \emph{close} to a symmetric state are affected too: there the low-$L$ features are small but nonzero, so that the product basis becomes ill-conditioned, and the coefficients $c_{k}$ must vary steeply, or even diverge as the symmetric state is approached, which makes them hard to learn.

In Fig.~\ref{fig:angular_order_benchmark}a we probe this shortcoming with MACE~\cite{batatia2022mace,batatia2025foundation,batatia2025design} on the on-site Hamiltonian blocks of rattled FCC-Au, whose environments are close to octahedral, with $\lambda_{\text{max}}^{\text{target}} = 6$~\cite{qian2026equivariant}. Each model is fitted to a single 64-atom structure and tested on 99 unseen ones, so that what it generalizes reflects what its basis can represent (Sec.~\ref{sec:si_fccau}).
Blocks with $\lambda > L_{\text{max}} \times \nu$ cannot be fitted at all, and those reached only through the product basis, $L_{\text{max}} < \lambda \le L_{\text{max}} \times \nu$, are fitted but often generalize no better than predicting zero.
Only blocks with $\lambda \le L_{\text{max}}$ are both fitted and generalized, and a basis-complete model does so even when reduced to the size of the smallest PET model (Sec.~\ref{sec:si_fccau_timings}).
Truncation is therefore not a viable strategy for a model that aims to be generally applicable, and a complete expansion carries the steep cost increase in $L_{\text{max}}$ discussed above.

There is a further trade-off associated with inversion symmetry. The geometric inputs of an equivariant model, the interatomic displacements and their spherical harmonics, are all proper, so that pseudotensorial features can only be formed by coupling at least two of them.
A $(1, -1)$ target, such as the antisymmetric part of the NMR shielding tensor, requires the model either to carry pseudotensorial channels alongside the proper ones, or to construct the target as a product of two learned tensors of matching parity~\cite{benmahmoud2025nmr}.
Either way, the antisymmetric block is consistently the hardest of the shielding tensor to learn~\cite{benmahmoud2025nmr,bornes2026nmr}.
In their naive form, $\mathrm{O}(3)$-equivariant networks must therefore carry separate proper and improper representations, making their tensor products four times more expensive than $\mathrm{SO}(3)$ ones, and in practice they often drop the improper representations altogether~\cite{batzner2022nequip,batatia2022mace,musaelian2023allegro}, potentially at the expense of completeness~\cite{bigi2026space}.
Some architectures instead enforce only $\mathrm{SO}(3)$ equivariance and learn inversion symmetry by augmentation~\cite{frank2022so3krates,fu2025esen,bigi2026space}, a partial step toward the fully O(3)-unconstrained approach taken here.

\subsection{Unconstrained GNNs}

Unconstrained GNNs take a different approach. PET~\cite{pozdnyakov2023pet,mazitov2025petmad} contains no explicit $\mathrm{O}(3)$ algebra: its backbone is built on edge-to-edge attention, in which the edges from an atom to all of its neighbors attend to each other, and the required symmetries are learned in training via data augmentation.
Backbone features are therefore generated at a cost independent of angular order, with no $L_{\text{max}}$ to truncate and no target irrep excluded by construction, at the price of satisfying Eq.~\eqref{eq:equivariance} only approximately.
There is also no explicit treatment of inversion symmetry.
PET's inputs are proper vectors too, so it \emph{must} also build pseudotensorial character by combining them, which it does through the products of edge features implicit in its edge-to-edge attention~\cite{domina2026unconstrained}, without any explicit coupling or channels of fixed angular order or parity.
These characters develop later in training than proper ones and remain weaker~\cite{domina2026unconstrained}, a point we return to in Sec.~\ref{sec:improving_equi}.

The symmetry of a target is only selected in the last layers of the model.
In PET-based MLIPs, the backbone features $\textbf{h}_{i(j)}$ of each node $i$ or edge $ij$ are mapped onto a target $t$ by a nonlinear head $\phi^{(t)}$, whose outputs are the last-layer features (LLF), and a linear readout for each irrep block $\alpha$:
\begin{equation}
    \tilde{y}^{(t, \alpha)}_{i(j)} = W^{(t, \alpha)} \phi^{(t)} \left( \textbf{h}_{i(j)} \right) + b^{(t, \alpha)},
    \label{eq:per_target_head}
\end{equation}
with edge predictions summed over neighbors $j$, per-system outputs additionally over atoms $i$, and a bias $b^{(t, \alpha)}$ for every irrep, which, as PET is unconstrained, is not forced to vanish for the non-invariant ones, as it would be in an equivariant model.
With this \textit{per-target} head, the LLF develop under augmentation a spectrum of $\mathrm{O}(3)$ characters far richer than the target requires, from which the linear readouts, unable to create new characters, must filter the one they are defined for~\cite{domina2026unconstrained}.
The burden is mild for the PES, whose targets span at most up to $\lambda = 2$ (when stress and forces are predicted directly instead of as energy gradients), but grows for a target of many irreps, all read out from one shared, mixed-character LLF vector.

We therefore make three modifications to the heads and readouts of PET, leaving the backbone and its cost untouched.
First, the node and edge head widths are set independently, each matched to the features it acts on, so that the contraction or expansion of the backbone features into the LLF can be controlled. This matters most for high-dimensional targets such as density expansions, where the expressivity can be limited by the width of the model.
Second, we introduce \textit{per-block} heads, conditioned on each irrep block $\alpha$ of the target $t$,
\begin{equation}
    \tilde{y}^{(t, \alpha)}_{i(j)} = W^{(t, \alpha)} \phi^{(t, \alpha)} \left( \textbf{h}_{i(j)} \right) + b^{(t, \alpha)},
    \label{eq:per_block_head}
\end{equation}
so that each LLF can specialize to a single character $(\lambda, \sigma)$ and its readout need only filter within it.
Third, for targets on an atom-centered basis, where each atomic type $Z$ carries its own basis functions, the readout is conditioned on the type of the central atom, $W^{(t, \alpha, Z_i)}$.
Solid-harmonic geometry embeddings, which we found necessary for isolated high-order targets~\cite{domina2026unconstrained}, make little difference for the targets considered here, which span all orders from $\lambda = 0$ upwards, and we use the standard embeddings throughout (see Sec.~\ref{sec:si_geometry_embeddings}).

Returning to Fig.~\ref{fig:angular_order_benchmark}, PET models of matched parameter count both fit and generalize every block, with errors that are nearly uniform across angular orders (panel b). The one exception is the smallest per-target model, which cannot learn the highest $\lambda$, an issue resolved by per-block heads on the same backbone. Every PET model is also cheaper at inference than the basis-complete MACE models: the smallest is 1.7 times faster than the MACE model of the same size with the fused cuEquivariance kernels (Fig.~\ref{fig:si_angular_order_timings}). No hyperparameter of PET is tied to the angular order of the target, and per-block heads add little to the cost of the backbone, so that the main design decision is the backbone size, a trade-off between accuracy, speed and memory.

\section{Learning equivariant atomistic properties with unconstrained GNNs}
\label{sec:learning}

\subsection{A simple but versatile architecture} \label{sec:size_ablations}

We now demonstrate that an unconstrained architecture can be used for learning general equivariant atomistic properties, with the same backbone across all targets. All models are implemented in and trained with \texttt{metatrain}, part of the \texttt{metatensor} and \texttt{metatomic} ecosystem~\cite{bigi2026metatomic}, with more details on the specific model and training hyperparameters provided in Sec.~\ref{sec:si_training} of the SI.

First, we define a set of dataset-agnostic backbone architectures, defined independently of the target symmetries, that vary under simple heuristics.
These are labeled by their size, \texttt{PET-XS} (2.8 M parameters), \texttt{PET-S} (11.2 M), \texttt{PET-M} (25.3 M), and \texttt{PET-L} (64.9 M), and obtained by linearly increasing the edge feature dimension as $d_{\text{PET}} \in \{128, 256, 384, 512\}$, with an additional transformer layer at size \texttt{L} and other internal dimensions defined relationally, i.e., $d_{\text{node}} = d_{\text{ff}} = 2 \times d_{\text{PET}}$. A fixed cutoff of 5\,\AA{} with two message passing layers is used for all backbone sizes.
Each backbone is combined with either per-target (Eq.~\eqref{eq:per_target_head}) or per-block (Eq.~\eqref{eq:per_block_head}) heads, and with target-dependent linear readouts. Combined, the head and readout modules are relatively lightweight compared to the backbone and make up a small fraction of the total inference cost.

Then, we choose three publicly available datasets that together span the challenges identified above.
Unless stated otherwise, all backbone sizes are trained with both head types.
a) SPICE-$\alpha$~\cite{gonnheimer2026macemdp} is a large dataset of 1.66~M molecular conformers with dipoles and polarizabilities, global targets that span the $(1, +1)$ and $(0, +1) \oplus (2, +1)$ irreps respectively, and are jointly learned. Dipoles are predicted as a sum of atomic vector contributions, $\boldsymbol{\mu} = \sum_i \boldsymbol{\mu}_i$. An alternative formulation is to add scalar atomic partial charges displaced to the atomic positions, $\sum_i q_i \textbf{r}_i$, which can describe non-local charge transfer that local atomic dipoles cannot~\cite{veit2020muml,unke2019physnet,gonnheimer2026macemdp,stark2026simultaneous}. We train backbone sizes up to \texttt{PET-M}.
b) ShiftML-molecules~\cite{kellner2026shiftml4} contains NMR chemical shielding tensors of 56\,085 closed-shell isolated molecules, computed with gauge-including atomic orbitals (GIAO) at the PBE0/cc-pVTZ level~\cite{ditchfield1974giao,wolinski1990giao}, a local target defined per atom that spans $(0, +1) \oplus (1, -1) \oplus (2, +1)$ and so includes a pseudotensorial block. The molecules are extracted whole from a database of organic crystals sampled from the Cambridge Structural Database~\cite{cordova2022shiftml2,kellner2026shiftml4}, and the training splits of the crystal dataset are carried over to the molecules extracted from it. While all nuclei are simultaneously learned, we report results for the most experimentally important $^{1}$H and $^{13}$C nuclei. We note that species-conditioned readouts add no benefit in the case of this dataset: species embeddings implicitly contained within the LLF carry enough information---in the absence of an explicit atomic basis---to map \textit{via} shared readouts onto the output space.
c) SCFBench~\cite{liu2026scfbench} contains the coefficients of the electron density expanded on several atom-centered auxiliary bases, a local target of proper irreps with angular order $\lambda \le 4$, with a number of radial basis functions that depends on the atomic type and on the basis set.

Figure~\ref{fig:equivariance_overview}a,b shows two contributions to the error of each target, with per-irrep values given in the SI (Fig.~\ref{fig:si_size_ablation_irreps}): the RMSE of the $\mathrm{O}(3)$-symmetrized prediction, which measures the accuracy of the model independently of its symmetry breaking, and the equivariance RMSE, the per-irrep equivariance error ($A_{\alpha}$ of Ref.~\onlinecite{domina2026unconstrained}), given by the spread over the group orbit of the back-transformed prediction $\rho^{-1}(g) f(gA)$.
Errors on equivariant quantities are reported as RMSEs throughout, as the mean absolute error (MAE) is not invariant to rotations for $\lambda > 0$.
Here we compare models by accuracy and equivariance only. The details of their inference cost and memory footprint, as well as their comparison with external references, are deferred to Sec.~\ref{sec:fast_expressive}.

Scaling the backbone of per-block models systematically improves both the symmetrized and the equivariance errors together for every dataset and target (Fig.~\ref{fig:equivariance_overview}a), with isolated exceptions for single irreps, such as the equivariance error of the $(1, -1)$ $^{13}$C block from \texttt{PET-XS} to \texttt{PET-S} (Fig.~\ref{fig:si_size_ablation_irreps}).
The gains are largest for SCFBench, by far the highest-dimensional target we consider, where a wider model is most likely to alleviate any rank deficiency in the mapping from backbone features to outputs.
The choice of head is a secondary effect (Fig.~\ref{fig:equivariance_overview}b).
Per-block heads leave the accuracy unchanged or improve it slightly, and their effect on the equivariance error grows with the size of the backbone: neutral or slightly adverse for the smallest models of the shieldings, it becomes a consistent reduction for the shieldings and the density from \texttt{PET-M} on.
For SPICE-$\alpha$, where per-target and per-block head types differ only by a single head, the choice makes little difference.
For the low-order targets of SPICE-$\alpha$ and ShiftML the equivariance error is a minor contribution to the total.
For SCFBench, however, its share grows with angular order. From \texttt{PET-XS} to \texttt{PET-L} the equivariance error of the $\lambda = 3$ and 4 blocks falls three- to fourfold, but their symmetrized error falls almost as much, so that the former remains at half or more of the latter (Fig.~\ref{fig:si_size_ablation_irreps}).

\begin{figure*}[t]
    \centering
    \includegraphics[width=\linewidth]{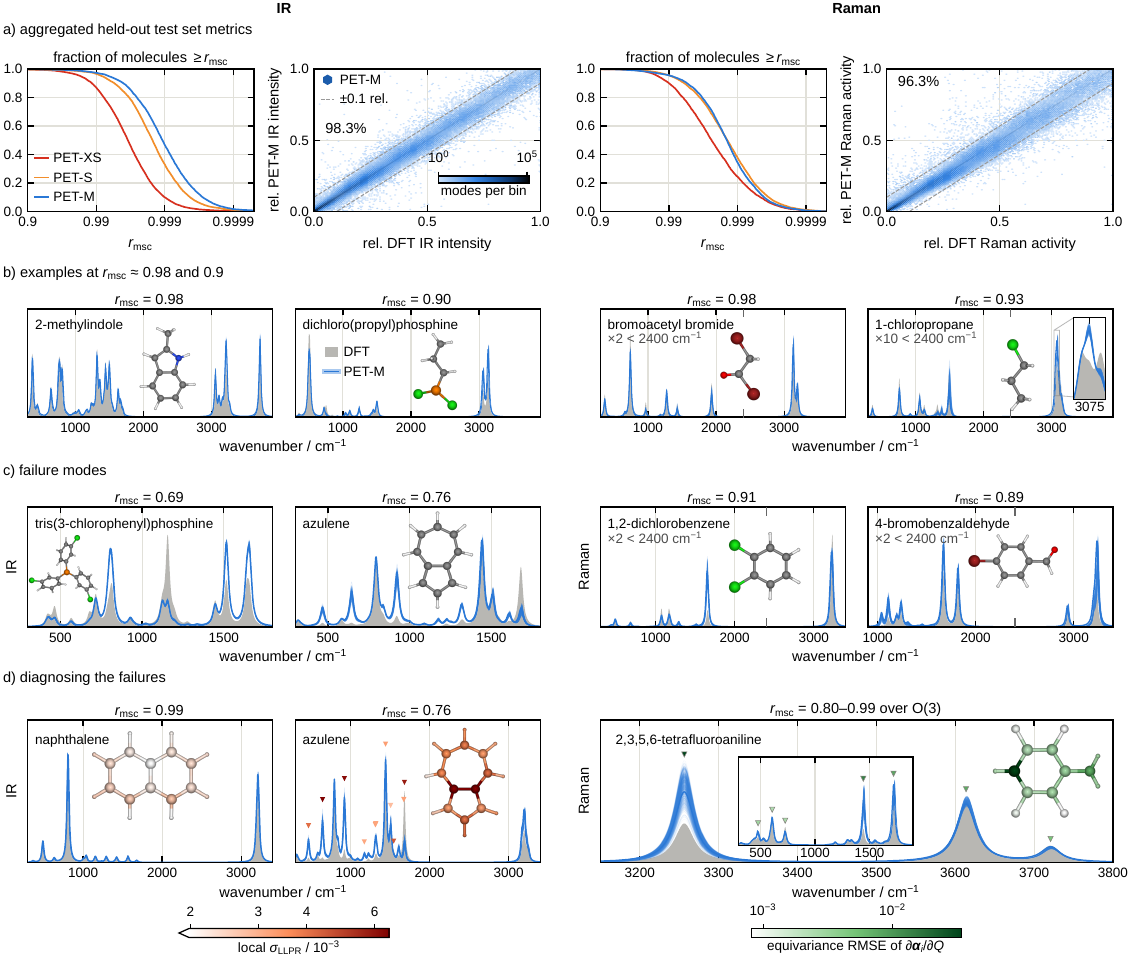}
    \caption{
        IR and Raman spectra of the SPICE-$\alpha$ PET models for the 7184 molecules of IR-R-7193 scored in the benchmark (see text), at the DFT normal modes and broadened with a FWHM of $30\,\mathrm{cm^{-1}}$. DFT is in gray and \texttt{PET-M}, in the reference orientation of each molecule, in blue; in b)--d), thin light-blue lines show the \texttt{PET-M} spectra over the 760 operations of an $\mathrm{O}(3)$ quadrature (degree 9, with inversions), so that their spread is the equivariance error of the spectrum. Agreement is measured by the match score $r_{\mathrm{msc}}$, the squared cosine similarity of the broadened DFT and PET spectra. Where indicated, intensities below $2400\,\mathrm{cm^{-1}}$ are scaled for visibility. a) Fraction of molecules with match score at least $r_{\mathrm{msc}}$ for the XS, S and M models, on a logarithmic scale in $1 - r_{\mathrm{msc}}$; and parity of the \texttt{PET-M} intensities, normalized to each molecule's strongest mode, with the fraction of strong modes (above 0.1) predicted within $\pm 0.1$. b) Spectra at $r_{\mathrm{msc}} \approx 0.98$ and 0.9, among the lowest 2\% and 0.2\% of the benchmark, scaled to the norm of the DFT spectrum; the inset magnifies the C--H stretch band of 1-chloropropane. c) Some of the lowest-scoring molecules, with frequency ranges cut to the relevant peaks. d) Left: naphthalene and azulene, with atoms colored by their local LLPR uncertainty, and triangles, colored on the same scale, marking the modes along which a ring-fusion carbon of azulene is more uncertain than any atom of naphthalene. Right: Raman spectrum of 2,3,5,6-tetrafluoroaniline, whose $r_{\mathrm{msc}}$ ranges from 0.80 to 0.99 over the quadrature. Atoms are colored by the equivariance RMSE of their contribution $\partial\boldsymbol{\alpha}_i/\partial Q$ (largest over modes), and triangles mark the modes whose DFT or PET activity exceeds 5\% of the largest DFT activity, colored by the largest atomic equivariance RMSE along them; the inset magnifies the fingerprint region.
    }
    \label{fig:spectra}
\end{figure*}

\subsection{Improving equivariance errors}
\label{sec:improving_equi}

Whatever its share of the total error, the equivariance error of a single model may still be too large for a downstream application, and we now consider two strategies that target it directly. We take the models of $^{13}$C chemical shieldings as an example, as these are the models validated on chemical shifts in Sec.~\ref{sec:validation_nmr} (Fig.~\ref{fig:equivariance_overview}c,d; $^{1}$H in Fig.~\ref{fig:si_ensemble_equivariance}).

The first is the choice of head type, which we analyze with the $\mathrm{O}(3)$-character decomposition (the ``B-metric'' of Ref.~\onlinecite{domina2026unconstrained}) applied to the LLF of \texttt{PET-L} (Fig.~\ref{fig:equivariance_overview}c, left).
With a per-target head a single set of LLF serves all three irreps of the shielding tensor, and its character is a compromise between them.
With per-block heads the character of each target irrep is more strongly expressed in the LLF of its own head, most markedly for the weakly expressed pseudovector.
This does not guarantee character-pure LLF---e.g., those of the $(1, -1)$ head remain dominated by proper characters that the readout must still filter out---but it removes part of the burden from the readout, and the equivariance error falls for every irrep (Fig.~\ref{fig:equivariance_overview}c, right). For \texttt{PET-M} and \texttt{PET-L}, the magnitude of this effect (13 to 25\%) is larger than the seed-to-seed variation (7 to 13\%) between ensemble members, introduced and discussed below.
Post-hoc linear purification~\cite{domina2026unconstrained}, in which the readout weights are refit for each irrep and atomic type under a penalty of weight $\gamma$ on the equivariance error, lowers the equivariance error by only 2 to 11\% for both head types at unchanged total error (Sec.~\ref{sec:si_purification}). The remaining symmetry breaking therefore cannot be filtered out of the LLF linearly, and must be reduced in the features themselves.

The second is ensembling. Equivariance emerges systematically in deep ensembles as the committee grows~\cite{gerken2024emergent}, and ensembling was found to strongly suppress the rotational fluctuations of a single unconstrained shielding model, which were comparable to its total error~\cite{kellner2025shiftml3}.
In a deep ensemble~\cite{lakshminarayanan2017ensembles}, $N$ models trained from different random seeds provide $N$ independent samples of the symmetry-breaking part of the prediction, which averages out in their mean.
The equivariance RMSE accordingly closely follows the $N^{-1/2}$ trend for every irrep and backbone size, even at small $N$ (Fig.~\ref{fig:equivariance_overview}d, left).
The PET-based ShiftML4 architecture from Refs.~\onlinecite{kellner2025shiftml3,kellner2026shiftml4}, with an edge feature dimension of $d_{\text{PET}} = 192$ and trained on the same dataset, shows the same trends at generally higher errors, which we discuss in Sec.~\ref{sec:validation_nmr}.
In all cases, deep ensembling is a simple and systematic way to reduce the equivariance error, but an expensive one~\cite{schafer2026shallow}: the gain in accuracy saturates, while the training and inference costs grow essentially linearly with $N$ (Fig.~\ref{fig:equivariance_overview}d, right; see Sec.~\ref{sec:si_timing}), though overhead-sharing (e.g., batch preparation) reduces this in the small-$N$ regime.

Partial ensembles~\cite{kellner2024shallow}, whose members share weights from part of the backbone, reduce this cost at the expense of members that are no longer independent.
For \texttt{PET-L}, branching at the heads, with the whole backbone shared, brings almost no benefit (Fig.~\ref{fig:si_ensemble_equivariance}), whereas branching after the first GNN layer gives each member its own second layer, with seed variation entering through the initialization of its geometry embeddings~\cite{domina2026unconstrained} and the augmentations each member sees.
The $N^{-1/2}$ trend expected for independent samples no longer holds, but a systematic improvement in both equivariance and accuracy is recovered, at both lower training (see Sec.~\ref{sec:si_training_ensembles}) and inference cost than the deep ensemble.

\section{Physical validation}
\label{sec:validation}

Errors on a held-out test set measure how well a model reproduces its target, but not how those errors, nor the residual symmetry breaking, propagate to the quantities that are ultimately of interest, which are rarely the direct targets themselves.
Vibrational spectra depend on the derivatives of the dipole and polarizability along the normal modes, chemical shifts on the isotropic part of the shielding tensor relative to a reference, and a predicted electron density is only as useful as the electronic-structure calculation it initializes or replaces.
We therefore validate the models of Sec.~\ref{sec:learning} on typical downstream tasks: the IR and Raman spectra of molecules with the SPICE-$\alpha$ models, NMR chemical shifts with the ShiftML models, and the use of predicted densities as initial guesses for DFT with the SCFBench models.
We then compare all of them with the external references in terms of accuracy, inference cost and memory footprint.

\subsection{IR/Raman spectra}
\label{sec:validation_spectra}

First, we validate the per-block XS, S, and M PET models, trained on SPICE-$\alpha$, on the IR-R-7193 held-out~\cite{gonnheimer2026zenodo}. It contains $\omega$B97M-D3(BJ)/def2-TZVPPD IR and Raman spectra for 7193 molecules and is the benchmark used for MACE-MDP~\cite{gonnheimer2026macemdp}. Frequencies and normal modes are taken from the DFT reference, so every difference between PET and DFT spectra comes from the property models---those predicting $\boldsymbol{\mu}$ and $\boldsymbol{\alpha}$. Spectral intensities are computed via central finite-difference derivatives of the model predictions, with a displacement of $10^{-3}$\,\AA\,amu$^{1/2}$ along each normal-mode coordinate (see Sec.~\ref{sec:si_spectra_intensities}).
Spectra are Lorentzian-broadened with a FWHM of 30\,cm$^{-1}$, and agreement is quantified with the match score $r_\mathrm{msc}$, the squared cosine similarity of the broadened reference and predicted spectra~\cite{gonnheimer2026macemdp}.
On top of the 4 open-shell molecules excluded in Ref.~\onlinecite{gonnheimer2026macemdp}, we exclude 5 molecules with unphysical or incorrect reference intensities, leaving 7184 molecules. Of these, 47 have a reference geometry whose bond graph agrees neither with PubChem nor with the NIST structure of its CAS number; their spectra are consistent with the computed structure, so they are scored, but they are not used as examples (see Sec.~\ref{sec:si_spectra_exclusions}).

All three models reproduce the reference spectra well (Fig.~\ref{fig:spectra}a,b), with mean scores of 0.994--0.997. Over all modes, the mean absolute intensity error falls from 2.1 to 1.2\,km\,mol$^{-1}$ (IR) and from 2.7 to 2.0\,\AA$^4$\,amu$^{-1}$ (Raman) from XS to M (Table~\ref{tab:si_spectra}). The mean score, which is insensitive to the overall intensity scale and close to saturation, does not separate S from M in either channel. For IR, the M model nonetheless has fewer poorly predicted spectra (197 against 249 with $r_\mathrm{msc} < 0.99$, Fig.~\ref{fig:spectra}a), while for Raman only the mode-level error separates the two. Evaluated at the same DFT normal modes, MACE-MDP has a mean score of 0.996 in both channels and mode-level errors between those of the XS and S models, although its total intensities are less biased (Table~\ref{tab:si_spectra}). The 0.88--0.97 reported for it in Ref.~\onlinecite{gonnheimer2026macemdp} also include the errors of the normal modes of its underlying MLIP.

The remaining failures (Fig.~\ref{fig:spectra}c) are concentrated on chemistry that is rare in SPICE-$\alpha$, such as triaryl phosphines, of which SPICE-$\alpha$ contains a single example against four in IR-R-7193, which suggests training-set coverage as the cause. 
We test this with the last-layer prediction rigidity (LLPR)~\cite{bigi2024rigidity}, which treats the last layer of the model as a linear regression on the frozen last-layer features, and the uncertainty it implies is large when the features of a new structure point in directions that the training set rarely samples. Since an intensity is the derivative of $\boldsymbol{\mu}$ or $\boldsymbol{\alpha}$ along a normal mode, which often involves only a few atoms in a local environment, we use the local version of LLPR~\cite{chong2023robustness} and evaluate it on the features of each atom, at the reference geometry and along each normal mode, flagging each molecule by its most uncertain atom (Sec.~\ref{sec:si_llpr}). Figure~\ref{fig:spectra}d (left) compares azulene with its isomer naphthalene, which PET predicts well. The ring-fusion carbons of azulene are more than twice as uncertain as any atom of naphthalene, and the modes along which they move are those with the largest errors. Over the whole benchmark, the 5\% most flagged molecules contain 29 of the 32 IR and 24 of the 29 Raman spectra with $r_\mathrm{msc} < 0.95$. Since such failures are rare, most of the flagged molecules are still predicted well. The LLPR therefore points to the spectra that should not be fully trusted, although the LLPR-derived uncertainties are not quantitatively correlated with the actual magnitude of the errors.

Finally, we tested whether the failures come from the approximate equivariance of PET. As a screening, we predicted every spectrum with $r_\mathrm{msc} < 0.98$ outside the mislabeled geometries in 24 random orientations of the molecule, and found that these spectra depend much more on orientation than well-predicted ones. However, averaging the derivatives over these orientations only raises $r_\mathrm{msc}$ appreciably for 4-bromobenzaldehyde (from 0.89 to 0.96). Within the failing spectra, the orientation spread also tends to mark the mispredicted peaks: the strong peak furthest from DFT is among the three most orientation-sensitive ones in 44 of 61 IR and 105 of 126 Raman spectra, against 14 and 57 expected by chance (Sec.~\ref{sec:si_spectra_orientation}).
For the molecules of Fig.~\ref{fig:spectra}, we repeated the analysis on a converged $\mathrm{O}(3)$ quadrature, including inversions, which confirms these findings. In 2,3,5,6-tetrafluoroaniline, the Raman spectrum most sensitive to orientation ($r_\mathrm{msc}$ from 0.80 to 0.99, Fig.~\ref{fig:spectra}d, right), the spread comes almost entirely from the contribution of the carbon bonded to the only ring hydrogen, yet the orientation-averaged spectrum is no closer to the reference.
Symmetry breaking is thus likely a symptom, like the LLPR, of the environments that the model has not learned well, rather than the cause of the failures, which lies in the coverage of the training set. Ensembling (Sec.~\ref{sec:improving_equi}) would reduce the equivariance errors of these nuclei, but since exact orientation averaging does not improve their spectra, any gain would have to come from averaging the errors of independent members.

\subsection{NMR chemical shieldings}\label{sec:validation_nmr}

\begin{figure}[tp]
    \centering
    \includegraphics[width=\linewidth]{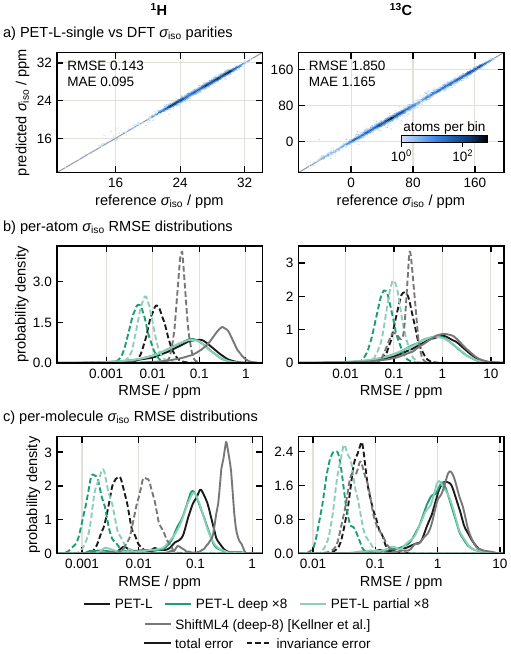}
    \caption{
    Validation of the isotropic chemical shielding $\sigma_{\text{iso}}$ of $^{1}$H and $^{13}$C nuclei on the ShiftML-molecules test set. Quantities are evaluated in the Cartesian basis, with $\sigma_{\text{iso}}$ taken as a third of the trace of the predicted shielding tensor.
    a) Parity of the per-atom predictions of the single \texttt{PET-L} model against the reference, in the original orientation of the 990 structures of the equivariance subset, hex-binned and colored by the number of atoms.
    b) Distributions over all 4105 test structures of the per-atom total error (solid) and invariance error (dashed) for \texttt{PET-L}, its 8-member deep and partial ensembles, and the 8-member ShiftML4 ensemble~\cite{kellner2025shiftml3,kellner2026shiftml4}. Both are computed from predictions on a set of 76 orientations of each structure (Sec.~\ref{sec:si_quadrature}): the total error is the absolute error of the orientation-averaged prediction, and the invariance error the root-mean-square variation of the prediction over O(3).
    c) The same for the RMSE over all nuclei of the given type in a molecule, averaged over the orientations for the total error, and with its root-mean-square variation over O(3) as the invariance error.
    }
    \label{fig:isotropics}
\end{figure}

\begin{figure*}[tp]
    \centering
    \includegraphics[width=\textwidth]{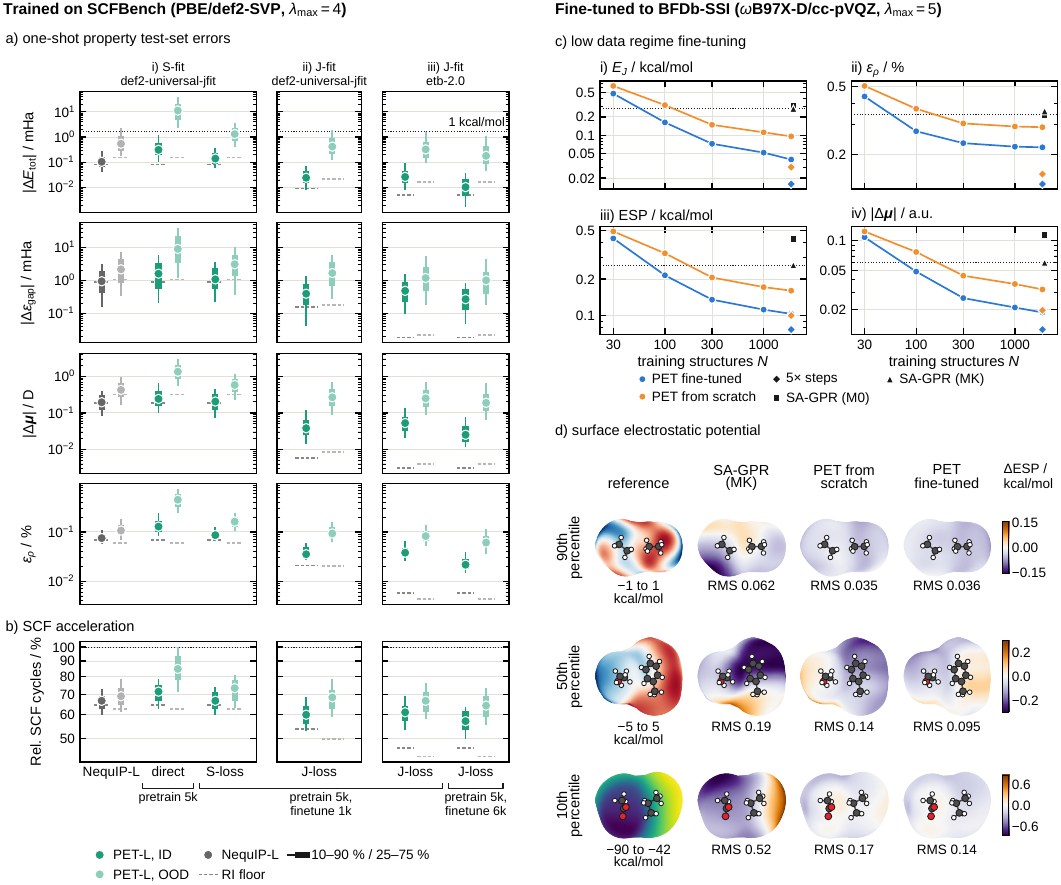}
    \caption{
    Density models trained on SCFBench (left) and fine-tuned to BFDb-SSI (right). a), b) \texttt{PET-L} at the PBE/def2-SVP level on 250 structures from each of the SCFBench in-distribution (ID) and out-of-distribution (OOD) test sets: i) S-fitted \texttt{def2-universal-jfit} coefficients, for NequIP-L~\cite{liu2026scfbench} and for \texttt{PET-L} trained on the coefficients and fine-tuned with the S density loss; ii) J-fitted coefficients with the J density loss; iii) the same in the larger \texttt{etb-2.0} basis, at equal and at extended training. a) One-shot errors after a single Fock matrix build and diagonalization; b) SCF cycles needed for convergence relative to the \texttt{minao} guess. Markers give medians and bars the 25--75\% and 10--90\% ranges; dashed lines give the RI floor. c), d) BFDb-SSI~\cite{briling2021metrics}, against the J-fitted reference density on the 322 test structures, with the electron-number correction (Sec.~\ref{sec:si_electron_correction}). c) Learning curves of \texttt{PET-L} fine-tuned from SCFBench or trained from scratch, after $2\times10^{4}$ optimizer steps (circles), and at $1\times10^{5}$ steps for $N = 1900$ (diamonds), against SA-GPR trained on all 2000 structures, without (M0) and with (MK) the electron-number constraint in training, both evaluated after the same correction. d) Surface ESP for test structures at the 90th, 50th and 10th percentile of the accuracy of SA-GPR (MK), with surface RMS errors in kcal/mol.
    }
    \label{fig:density_validation}
\end{figure*}

Next, we turn to the models trained on ShiftML-molecules, comparing the single \texttt{PET-L} model against its 8-member deep and partial ensembles from Sec.~\ref{sec:improving_equi}, with the released 8-member deep ensemble of the ShiftML4 architecture of Refs.~\onlinecite{kellner2025shiftml3,kellner2026shiftml4} as a reference, trained by its authors under their own protocol.
All metrics exclude 32 test structures, the conformers of six molecules with extreme reference $^{15}$N shieldings (Sec.~\ref{sec:si_nmr}).
The observable of interest is the isotropic chemical shift, $\delta$, which follows from the isotropic shielding $\sigma_{\text{iso}} = \mathrm{Tr}(\boldsymbol{\sigma}) / 3$ in the Cartesian basis.
Only invariant $(0, +1)$ terms of the model predictions contribute to it, so that the equivariance error reduces to an invariance error.
The shift is obtained through experimental calibration of the DFT shieldings, and as the two are linearly related we focus only on the latter.

Per atom, the ensembles lower the total error of the single \texttt{PET-L} model moderately, while their invariance errors separate clearly, falling from the single model to the partial and then the deep ensemble (Fig.~\ref{fig:isotropics}b).
Relative to ShiftML4, the total error of \texttt{PET-L} is markedly lower for $^{1}$H, but comparable for $^{13}$C. Part of the difference for $^{1}$H lies in the training protocol: ShiftML4 normalizes all targets by a single scale pooled over all nuclei, which strongly de-weights $^{1}$H in the loss, and retraining PET with the same convention increases its $^{1}$H error two- to threefold while raising that of $^{13}$C by at most 20\% (Fig.~\ref{fig:si_sml4_baseline_ablation}).

In NMR crystallography, where candidate structures are ranked by comparing computed and measured chemical shifts~\cite{salager2010powder,hodgkinson2020nmrcrystallography}, the quantity of interest is not the per-atom RMSE but rather the RMSE after aggregation over all nuclei of a given type in a molecule, as this matches more closely what experiments actually probe~\cite{kellner2025shiftml3,kellner2026shiftml4}.
Aggregation leaves the total errors of the same order, but the invariance errors drop relative to their per-atom counterparts (Fig.~\ref{fig:isotropics}c), an effect similar to that of ensembling, which likewise averages over per-atom predictions.
Deep ensembling improves the invariance error over a single model in all cases, and partial ensembling offers a trade-off between the two.
On this downstream quantity, however, the single \texttt{PET-L} model already achieves total and invariance errors that match or improve on those of the ShiftML4 ensemble, and ensembling remains available where a lower invariance error is still needed.
The motivating application is structure determination in molecular solids, whose reference shieldings are computed with the periodic gauge-including projector augmented wave (GIPAW) method~\cite{pickard2001gipaw,yates2007gipaw}, whereas the models here are trained and validated on isolated molecules, against DFT rather than measured shifts; the transfer of this analysis to the condensed phase, and the comparison with experiment that it enables, are left to future work.

\subsection{Electron densities}\label{sec:validation_density}

\subsubsection{Improving performance through training protocol}

Finally, we turn to the density models trained on SCFBench.
Among the representations used to learn the electron density, which include its values at real-space grid or probe points~\cite{jorgensen2022deepdft,focassio2023jacobi}, basis functions at learned positions~\cite{elsborg2025electra} and plane-wave expansions~\cite{elsborg2026electrafi}, we use the coefficients of an atom-centered auxiliary-basis expansion~\cite{grisafi2019transfer,fabrizio2019noncovalent,lewis2021condensed,grisafi2023electronic,rackers2023cracking,fu2024recipe,qiao2022informing,liu2026scfbench,lou2026moire}: these are compact, carry the per-atom equivariant structure of Sec.~\ref{sec:symmetries}, and live on an atomic basis in the same way as the shielding tensor, so that the same architecture serves all three datasets.
The coefficients $\textbf{c}^{\text{aux}}_A = \{ c^{\lambda \sigma}_{i \mu n} \}$ of a structure $A$ define a density through their expansion on the auxiliary basis,
\begin{equation}
    \rho^{\text{aux}}_A(\textbf{r}) = \sum_{i \lambda \mu n} c^{\lambda \sigma}_{i \mu n} \, \phi^{\lambda}_{i \mu n}(\textbf{r}),
    \label{eq:ri_density}
\end{equation}
where $\phi^{\lambda}_{i \mu n}$ is a basis function centered on atom $i$. Each coefficient is the component $\mu$ of a proper irrep $(\lambda, \sigma = +1)$, with $n$ indexing the radial function.
We denote by $\tilde{\textbf{c}}^{\text{aux}}_A$ the coefficients predicted by the model, and by $\tilde{\rho}^{\text{aux}}_A$ the density they define.
We assess this density in the two ways it would be used in practice.
In a \textit{one-shot} calculation, the Fock matrix is built from $\tilde{\rho}^{\text{aux}}_A$ and diagonalized once, with no SCF iterations, yielding a density matrix $\textbf{D}_0$ that grants access to all ground-state electronic properties. The HOMO--LUMO gap can be computed from the eigenvalues of the Fock matrix built from $\textbf{D}_0$.
As $\textbf{D}_0$ is built from the occupied orbitals, it holds the correct number of electrons by construction, so that, unlike the evaluations below that use the density in real space, these properties need no electron-number correction.
For larger systems, this Fock matrix build and diagonalization themselves become expensive; computation of observables directly from the RI coefficients or approaches that stitch local predictions into the density matrix of the whole system~\cite{stishenko2026scf} offer ways around this, though we focus on one-shot properties here.
We compare against the converged SCF solution the total energy, HOMO--LUMO gap, dipole moment, and real-space density, the latter through the normalized $L^1$ error
\begin{equation}
    \epsilon_\rho = \frac{100}{N_e} \times \int |\tilde{\rho}(\textbf{r}) - \rho(\textbf{r})| \, d\textbf{r} \ ,
\end{equation}
in percent, where $N_e = \int \rho(\textbf{r}) \, d\textbf{r}$ is the number of electrons.
Alternatively, $\textbf{D}_0$ is used as the initial guess of a full SCF calculation with PySCF~\cite{sun2020pyscf}, converged to $10^{-10}$~Ha, and we report the number of cycles relative to that from the \texttt{minao} guess, the PySCF default, which we re-run with identical settings. The count excludes the single Fock matrix build that yields $\textbf{D}_0$, beside which the inference of the model is negligible, and other standard guesses, such as the superposition of atomic potentials~\cite{lehtola2019guess}, would shift the baseline.
All calculations are performed at the PBE/def2-SVP level on 250 structures from each of the in-distribution (ID) and out-of-distribution (OOD) test sets of Ref.~\onlinecite{liu2026scfbench}, with the results shown in Fig.~\ref{fig:density_validation}a,b.
Alongside each model we report the same metrics for the reference coefficients it was trained on, which set the accuracy attainable for the given auxiliary basis (the `RI floor').
As an external reference we evaluate the NequIP-L model released with SCFBench~\cite{liu2026scfbench} through the same pipeline.
We then improve PET's performance in three steps that change how the learning problem is posed, keeping the backbone architecture untouched: the loss (column i), the fitting metric of the reference coefficients (column ii), and the auxiliary basis (column iii).

In Sec.~\ref{sec:size_ablations} the coefficients were treated as a direct learning target, with every component weighted according to its relative scale between radial basis functions (Sec.~\ref{sec:si_scales}).
The quantity of physical interest is however the density, and an error in the coefficients translates into an error in the density through the metric of the auxiliary basis, which is not orthogonal.
The more appropriate choice is therefore a loss that minimizes the error in the density itself~\cite{grisafi2019transfer,fabrizio2019noncovalent,lewis2021condensed,grisafi2023electronic,song2026selfconsistent},
\begin{equation}
    \mathcal{L}_A = \left( \Delta \textbf{c}^{\text{aux}}_A \right)^{T} \textbf{M}_A \, \Delta \textbf{c}^{\text{aux}}_A,
    \label{eq:metric_loss}
\end{equation}
where $\Delta \textbf{c}^{\text{aux}}_A = \tilde{\textbf{c}}^{\text{aux}}_A - \textbf{c}^{\text{aux}}_A$ is the difference between predicted and reference coefficients and $\textbf{M}_A$ is a metric matrix between the basis functions of structure $A$, in the simplest case their overlap $\textbf{S}_A$.
Starting from the \texttt{PET-L} model of Sec.~\ref{sec:size_ablations}, trained for 5000 epochs directly on the coefficients, a further 1000 epochs of fine-tuning with this loss improves every metric (Fig.~\ref{fig:density_validation}a,b, column i).
The effect is largest on transferability: on the OOD set, the median energy error falls by almost an order of magnitude to below chemical accuracy.
On the ID set the fine-tuned model approaches both NequIP-L and the RI floor, although NequIP-L remains the more transferable of the two. The PET models were, however, trained with a number of epochs, batch size and learning rate chosen for a consistent ablation across backbone sizes, not tuned for the largest model on this dataset (Table~\ref{tab:pet-training-hypers}). While the model could be optimized further, at this point it is the reference coefficients, rather than the model, that limit the accuracy---this is most evident for the dipole moment, for which both models sit at the RI floor itself.

The reference coefficients distributed with SCFBench are fitted by minimizing the squared residual of the density, while the alternative common in electronic-structure codes minimizes its Coulomb self-repulsion~\cite{whitten1973coulombic}:
\begin{align}
    \textbf{c}^{\text{aux}, S}_A &= \argmin_{\textbf{c}} \int d\textbf{r} \, \left| \Delta \rho_A(\textbf{r}) \right|^2 , \label{eq:s_fit} \\
    \textbf{c}^{\text{aux}, J}_A &= \argmin_{\textbf{c}} \iint d\textbf{r} \, d\textbf{r}' \, \frac{\Delta \rho_A(\textbf{r}) \, \Delta \rho_A(\textbf{r}')}{\left| \textbf{r} - \textbf{r}' \right|} , \label{eq:j_fit}
\end{align}
with $\Delta \rho_A = \rho_A - \rho^{\text{aux}}_A$ the residual of the expansion of Eq.~\eqref{eq:ri_density}, for trial coefficients $\textbf{c}$, with respect to the reference density $\rho_A$.
The two are referred to as S- and J-fitting after the overlap and Coulomb metrics that they involve~\cite{briling2021metrics}, and each defines the matching metric $\textbf{M}_A$ of Eq.~\eqref{eq:metric_loss}.
By minimizing the Coulomb self-repulsion of the residual, J-fitting leaves an error in the Hartree energy of the fitted density that is only second order in $\Delta \rho_A$, whereas S-fitting leaves a first-order term~\cite{dunlap2000robust,ren2012resolution}. The J-fitted density is also the one from which the Coulomb matrix is built in RI-J calculations, which makes it the natural target for SCF initial guesses.
With no change to the basis, moving from S- to J-fitted coefficients lowers the RI floor for every quantity, by a factor of seven to nine for the energy and by more for the dipole moment (dashed lines in Fig.~\ref{fig:density_validation}a,b, column ii).
The J-fitted densities are also neutral to within $6\times10^{-4}$~$e$, whereas the S-fitted ones carry a mean absolute charge of 0.02 to 0.08~$e$. This is a property of the Coulomb metric, which strongly penalizes any net charge of the residual, rather than of the electron-number constraint, which is imposed only on the \texttt{def2-universal-jfit} training set (Sec.~\ref{sec:si_ref_coeffs}).
A model trained on the J-fitted coefficients under the same protocol inherits much of this gain, in both accuracy and transferability. It improves on NequIP-L in every metric on both test sets, although only marginally for the SCF cycles of the OOD set. NequIP-L, however, was trained on the S-fitted coefficients, and we have no J-fitted reference for it.

The choice of auxiliary basis matters too.
The even-tempered \texttt{etb-2.0} basis has more than twice the number of radial functions per atom of \texttt{def2-universal-jfit}, and lowers the RI floor further, most of all for the gap (column iii).
Under the same training protocol and despite the lower floor, there is little difference in performance for the model in the larger basis until the fine-tuning is extended from 1000 to 6000 epochs, after the same 5000 epochs of pretraining (Sec.~\ref{sec:si_density_ft}). This yields the most accurate model by every metric, but one that remains further from its floor than the rest in every metric except the energy.
For a basis of this size the accuracy is limited by the model and its training instead of by the reference. Indeed, increasing to an even larger ETB basis would lower the RI floor still~\cite{liu2026scfbench}, at the expense of an increasingly ill-conditioned and high-dimensional learning problem.

Together, these three steps lower the median one-shot energy error of the same architecture by more than an order of magnitude on both test sets, taking the OOD set from well above chemical accuracy to well below it, and save around 40\% of the SCF cycles required from the \texttt{minao} guess, a saving of the same order as reported for equivariant density models of inorganic materials~\cite{koker2024charge3net}.
How the learning problem is posed, through the loss, the fitting metric, and the basis, matters at least as much as the capacity of the model that solves it.

How strongly the equivariance error of the predicted coefficients carries over to the one-shot observables depends on the observable (Sec.~\ref{sec:si_scfbench_o3}). Over 52 orientations of 50 molecules from each test set, the median spread of the one-shot energy of the J-fitted models is at most 0.04~mHa, at the level of the numerical noise of the integration grid. This is expected, as the total energy is stationary at the self-consistent density and so insensitive to its errors at first order. The dipole is instead linear in the density, and its median spread of 0.02 to 0.05~D, while below the model error on both test sets, reaches about half of it in distribution, where the J-fitted models are most accurate. As the model error decreases the equivariance error too decreases, but its relative share grows (Fig.~\ref{fig:si_scfbench_o3_variance}), so that for the most accurate density models it is no longer negligible.
In general, one-shot properties are complex nonlinear functions of the density matrix, and so depend on the predicted RI coefficients of all angular orders, inheriting their overall approximate equivariance.
Understanding how model errors propagate through electronic-structure surrogates, and the role of indirect learning~\cite{suman2025design}, is left to future work.

\begin{figure}[tp]
    \centering
    \includegraphics[width=\linewidth]{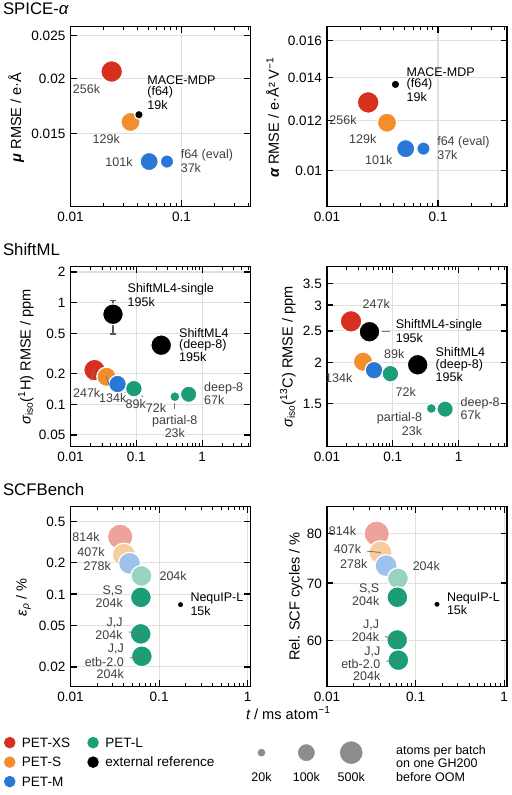}
    \caption{
        Accuracy against inference time per atom for per-block PET models of increasing backbone size (colored) and external references (black). Marker area gives the largest number of atoms that fits in a single batch on one NVIDIA GH200 before running out of memory, also written beside each point; inference times are measured at that batch size. Top: SPICE-$\alpha$ dipole $\boldsymbol{\mu}$ and polarizability $\boldsymbol{\alpha}$, RMSE over Cartesian components, against MACE-MDP in float64 as released, with \texttt{PET-M} also in float64. Middle: isotropic $^{1}$H and $^{13}$C shieldings, against the 8-member ShiftML4 ensemble~\cite{kellner2026shiftml4} and its single members (mean $\pm$ standard deviation), alongside 8-member deep and partial \texttt{PET-L} ensembles. Bottom: SCFBench one-shot density error $\epsilon_\rho$ and relative number of SCF cycles on the in-distribution test set, against NequIP-L. Lighter markers are models trained directly on the S-fitted coefficients; labeled \texttt{PET-L} variants are fine-tuned with a density loss and named by fitting metric and loss, S,S or J,J (Fig.~\ref{fig:density_validation}).
    }
    \label{fig:pareto}
\end{figure}

\subsubsection{Transfer-learning a new functional, basis, and chemistry}

The models trained on SCFBench are tied to the functional (PBE), orbital basis (def2-SVP) and auxiliary basis (\texttt{def2-universal-jfit} or \texttt{etb-2.0}) of their training data.
To test whether what they learn transfers to densities computed differently, we take the J-fitted, J-loss \texttt{PET-L} model and, with every parameter trainable, transfer-learn to the side-chain--side-chain dimers of the BioFragment Database~\cite{burns2017bfdb} as prepared in Ref.~\onlinecite{briling2021metrics}, and compare it with the same architecture trained from random weights under the same budget (Sec.~\ref{sec:si_finetuning}). Unlike for SCFBench, the observables are computed directly from the density of the predicted coefficients, after an electron-number correction discussed below, with no Fock matrix build.
Relative to SCFBench this changes the functional, the orbital basis, the auxiliary basis, which now reaches $\lambda = 5$, and the chemistry---around a fifth of the dimers carry a net charge---all of which the model has never seen.
The same dimers were the first benchmark for symmetry-adapted Gaussian process regression (SA-GPR) of the density~\cite{fabrizio2019noncovalent}, and the published predictions of Ref.~\onlinecite{briling2021metrics} provide a baseline that can be evaluated with exactly the metrics used for PET. The PET model trained from scratch is a modern baseline and isolates the effect of pretraining.

Both PET models improve on SA-GPR for every metric (Fig.~\ref{fig:density_validation}c): the Coulomb-metric error $E_J$, the real-space density error $\epsilon_\rho$, the error of the electrostatic potential (ESP) on the molecular isosurface of Ref.~\onlinecite{briling2021metrics}, and that of the dipole moment.
Starting from the SCFBench model, the fine-tuned model does so with 100 BFDb-SSI training structures and another 100 for validation and early stopping, a tenth of the 2000 on which SA-GPR is trained, while the model trained from scratch needs around 300.
Pretraining lowers the Coulomb-metric error by a factor of about two over training from scratch from 100 structures onwards, an advantage that survives a five-fold longer training.
Pretraining on PBE densities of small organic molecules thus transfers to a different functional, orbital basis and auxiliary basis, and lowers the error reached for a given amount of target data and training.

The error that remains is concentrated on the charged structures.
None of the PET models receives the total charge as an input, so that their predicted densities do not integrate to the correct number of electrons.
Following Ref.~\onlinecite{briling2021metrics} we correct this a posteriori for every model, which chiefly benefits the electrostatic potential (Sec.~\ref{sec:si_electron_correction}).
The error in the potential then remains small across the test set and improves on SA-GPR from the best to the worst cases (Fig.~\ref{fig:density_validation}d), while growing with the magnitude of the net charge, more steeply for SA-GPR than for PET.

Fine-tuning transfers in the same way to coupled-cluster densities.
On unrelaxed CCSD densities of QM9 molecules~\cite{li2026agentic}, ten training structures, with 80 more for validation and early stopping, bring the Coulomb-metric error of the pretrained models below that of B3LYP, and a few hundred bring it a factor of five below, consistent with earlier work in which coupled-cluster accuracy was reached from DFT densities with little high-level data~\cite{bogojeski2020cc}.
The gain is specific to the Coulomb metric in which the models are trained: the real-space density and the dipole moment improve only to about the B3LYP level, limited by the auxiliary basis and by the model respectively (Sec.~\ref{sec:si_cafes}).

\subsection{Fast and expressive}
\label{sec:fast_expressive}

Figure~\ref{fig:pareto} compares the PET models introduced above with external references, as released by their authors, in terms of accuracy, inference cost and memory footprint, and evaluated on the same hardware and under the same timing protocol as PET.
Across all three datasets PET traces a Pareto-optimal front.
We note, however, that MACE-MDP was evaluated as provided~\cite{gonnheimer2026macemdp}, i.e., without the cuEquivariance acceleration used in Fig.~\ref{fig:angular_order_benchmark}. PET, too, is timed without compilation.
For SPICE-$\alpha$, with each model evaluated in its training precision, \texttt{PET-S} (single precision) is faster than MACE-MDP (double) and more accurate for both targets.
When \texttt{PET-M} is evaluated in double precision it is slower but at no benefit to its accuracy: the precision is thus a design choice and here only serves as a point of comparison of memory footprints at matched precision.
For the isotropic shieldings, \texttt{PET-S} is both faster and more accurate than a single ShiftML4 model, and a single \texttt{PET-L} model is comparable to the 8-member ShiftML4 ensemble at less than half of its inference cost.
For SCFBench, the comparison depends on the target. Trained on the same S-fitted coefficients as NequIP-L, \texttt{PET-L} reaches comparable accuracy only after density-loss fine-tuning. When trained on J-fitted coefficients it improves on NequIP-L in density error and SCF acceleration at about a third of its cost per atom. As no J-fitted NequIP-L model exists, this gain reflects the better-posed target as much as the architecture.

The advantage of PET over equivariant models, however, is clearer when considering the memory footprint.
On a single GPU, \texttt{PET-L} fits 14 times as many atoms in one batch as NequIP-L, both in single precision. MACE-MDP is released in double precision, in which \texttt{PET-M} fits twice as many atoms. A smaller footprint allows correspondingly larger systems to be evaluated in a single pass, and ShiftML4, itself a PET-based architecture, shares this advantage.
For a given backbone size the batch capacity of PET differs between datasets, which reflects the density of neighbors within the fixed cutoff in the test molecules on which it is evaluated: the molecules of SCFBench are around half the size of those of SPICE-$\alpha$ and ShiftML, have fewer neighbors per atom, and so fit around three times as many atoms per batch.

\section{Discussion}\label{sec:discussion}

In this paper, we trained the unconstrained transformer-based GNN PET on a range of equivariant atomistic properties, and assessed its performance, trade-offs, and practical utility on downstream tasks.

With an architecture built around unconstrained edge-to-edge attention rather than CG tensor products, PET is able to learn in principle any O(3) symmetry from data, at an inference cost that is independent of the symmetries it targets. For a near-symmetric system, a small PET model generalizes up to $\lambda = 6$ from a single training structure. A truncated product basis fails to do so, and an equivariant GNN generalizes only once its features are complete up to the target angular order, at an inference cost that grows with that order.

The same PET backbone learns dipoles and polarizabilities, NMR shielding tensors including their pseudovector block (all $\lambda \le 2$), and the high-dimensional coefficients of the electron density ($\lambda \le 5$). Scaling backbone size by increasing feature widths improves accuracy and equivariance together. Per-block heads lower the equivariance error further for larger models, most clearly for the weakly expressed pseudotensorial block. Linear purification at fixed total error provides a marginal additional gain, and deep ensembles reduce the equivariance error as $N^{-1/2}$, with partial ensembles recovering part of this gain at lower cost.

On the observables derived from these models, the residual symmetry breaking is not what limits the accuracy. The SPICE-$\alpha$ models give IR and Raman spectra that match the reference closely, and their few failures involve chemistry that is rare in the training set, flagged both by the LLPR and by an orientation dependence of the prediction more than ten times that of well-predicted spectra. For the molecule-averaged chemical shieldings used in NMR crystallography, a single \texttt{PET-L} model matches or improves on the 8-member ShiftML4 ensemble.

For the electron density, how the learning problem is posed matters as much as the model. With a better choice of the loss function, the auxiliary basis and the metric under which the target coefficients are defined, all errors on one-shot properties decrease without changing the backbone. Pretrained density models transfer to a different functional, basis, and chemistry. Fine-tuned from SCFBench, PET improves on SA-GPR with a tenth of its BFDb-SSI data, and transfers to CCSD densities with fewer than a hundred CCSD calculations.

PET matches or improves on the released reference models at a lower inference cost for the dipoles, polarizabilities and shieldings, and for the density when trained under a more rigorous protocol. At the same precision, PET fits two to fourteen times as many atoms on a GPU as the equivariant models. 
Equivariance errors decrease systematically with model size and ensembling. Their correlation with model errors (found for the PES~\cite{domina2026unconstrained} and for rotational fluctuations of chemical shieldings~\cite{kellner2025shiftml3}) extends here to physical observables, so that the symmetry breaking of an unconstrained model can itself flag predictions that should not be trusted. Where exact symmetry is required, predictions can be averaged over an $\mathrm{O}(3)$ quadrature, at a cost proportional to its size~\cite{langer2024broken}.
The clearest limit is the electron density. Its equivariance error grows with angular order and, for $\lambda = 3$ and 4, falls with backbone size only about as fast as the symmetrized error, of which it remains half or more. For the most accurate models, the resulting spread of the one-shot dipole reaches about half the model error in distribution. As the derived properties are nonlinear functions of every angular order, how such errors propagate through electronic-structure surrogates, such as when indirect loss functions are involved~\cite{suman2025design}, deserves further study.

Several extensions follow directly. The density models receive no information about the total charge, and the remaining error on BFDb-SSI concentrates on the charged structures, so that charge-aware architectures~\cite{unke2021spookynet}, readouts with explicit partial charges~\cite{veit2020muml}, or training with the electron-number constraint built in~\cite{briling2021metrics} are natural next steps.
The auxiliary-basis machinery is not specific to the electron density: any scalar field expanded on an atom-centered basis, such as the spin density, the electrostatic potential or the local density of states, can be learned in the same way, and the electric field gradient tensor, a pure $(2, +1)$ target, requires no modification at all.
Electronic Hamiltonians, whose blocks reach the angular orders at which truncation fails and which present pseudotensorial targets, are an equally natural target. However, how best to manage the memory footprint for their off-site blocks (defined on pairs of atoms), and how best to enrich their learning of pseudotensorial terms, remain open challenges.

While inference cost and memory footprint will continue to improve for all models, equivariant or unconstrained, with specialized kernels and compilation methods~\cite{fu2026inference}, datasets will continue to become larger and more extensive in their coverage of chemical space and atomistic properties. Fast, expressive, and scalable architectures that are systematically improvable with size and training resources are likely to become the default for learning such targets and pushing the reach of ML surrogate models for atomistic simulation into new domains.
Unconstrained architectures are promising candidates for such tasks, learning O(3)-equivariance from data instead of enforcing it to unlock greater speed and accuracy, with a simple backbone whose cost is independent of the symmetries it targets.

\begin{acknowledgments}
The authors would like to acknowledge support by the following grants: the ERC Horizon 2020 Grant No. 101001890-FIAMMA (JWA,PF,MC,SS,PL); an Industrial Grant with Novartis, agreement No. 6480068733 (PP); an SNSF project, ID 200020\_214879 (MD,MK,MC,FB); NCCR MARVEL, a National Centre of Competence in Research, funded by the Swiss National Science Foundation, grant number 205602 (PF,SS,PL); the German Research Foundation (DFG) under Germany’s Excellence Strategy - EXC 2089/2 – 390776260 (PL).
Computation for this work relied on resources from the Swiss National Supercomputing Centre (CSCS) and the SwissAI program under the lp161, lp140, and ab010 projects, and the EPFL HPC platform (SCITAS).

AI tools have been used to assist with writing plotting and analysis scripts, and to improve the clarity and grammar of the manuscript.
\end{acknowledgments}

\section*{Data and code availability}

All models are trained with \texttt{metatrain}~\cite{bigi2026metatomic} on publicly available datasets: FCC-Au~\cite{qian2026equivariant}, SPICE-$\alpha$~\cite{gonnheimer2026macemdp}, ShiftML-molecules~\cite{kellner2026shiftml4}, SCFBench~\cite{liu2026scfbench}, BFDb-SSI~\cite{burns2017bfdb,briling2021metrics} and CAFES-QM9~\cite{li2026agentic}. The spectra are validated on IR-R-7193~\cite{gonnheimer2026zenodo}, and the reference models and SA-GPR predictions are those released with Refs.~\onlinecite{gonnheimer2026macemdp,kellner2026shiftml4,liu2026scfbench,briling2021metrics}.
Raw data and plotting scripts to reproduce the figures, model checkpoints, and input files will be provided in an open online data repository (DOI to be confirmed).

\section*{Conflicts of interest}

There are no conflicts to declare.

\clearpage
\onecolumngrid

\setcounter{section}{0}
\setcounter{equation}{0}
\setcounter{figure}{0}
\setcounter{table}{0}
\renewcommand{\thesection}{S\arabic{section}}
\renewcommand{\theequation}{S\arabic{equation}}
\renewcommand{\thefigure}{S\arabic{figure}}
\renewcommand{\thetable}{S\arabic{table}}
\makeatletter
\@ifpackageloaded{hyperref}{%
  \renewcommand{\theHsection}{S\arabic{section}}%
  \renewcommand{\theHequation}{S\arabic{equation}}%
  \renewcommand{\theHfigure}{S\arabic{figure}}%
  \renewcommand{\theHtable}{S\arabic{table}}%
}{}
\makeatother

\suppressfloats[t]
\begin{center}
    {\large \textbf{Supplementary Information}}\\[0.5em]
    \textbf{Unconstrained learning of equivariant atomistic properties}
\end{center}
\vspace{1em}

\section{Basis incompleteness}

\subsection{Beyond completely vanishing spherical harmonics}
\label{sec:si_dimensionalityanalysis}

Consider a basis for angular order $\lambda$ built from products of features of angular orders $l_1$ and $l_2$,
\begin{equation}
    \mathbf{B}^{\lambda}_i = \sum_{l_1 n_1, l_2 n_2} c_{l_1 n_1, l_2 n_2} \left( \mathbf{B}^{l_1}_{n_1} \ocg \mathbf{B}^{l_2}_{n_2} \right)_{\lambda} ,
    \label{eq:si_product_basis}
\end{equation}
where each $\mathbf{B}^{l}_{n}$ is a feature of angular order $l$, with $2l + 1$ components.
This construction fails whenever all features of order $l_1$ or $l_2$ vanish, as they do when symmetry forces the corresponding spherical expansion to zero.
Octahedral environments, for which it vanishes for $l = 1$, 2, 3 and 5, are the most familiar example (main text), but a vanishing expansion is only the extreme case of a more general problem.

The problem is best seen geometrically.
Just as three linearly independent vectors are needed to span three-dimensional space, $2\lambda + 1$ linearly independent tensors of order $\lambda$ are needed to span the $(2\lambda + 1)$-dimensional space of that order.
Given a complete basis for order $\lambda$, products of its $2\lambda + 1$ tensors, as in Eq.~\eqref{eq:si_product_basis}, reach orders up to $2\lambda$, and their $(2\lambda + 1)^2$ combinations are more than enough to provide the $4\lambda + 1$ independent tensors needed for order $2\lambda$.
If the expansion of order $\lambda$ vanishes, however, there are no tensors to multiply, and no products of order $2\lambda$ can be formed, although that space need not be empty, as the $l = 4$ expansion of the octahedron shows.
Between these extremes, symmetry can restrict the space of order $\lambda$ to fewer than $2\lambda + 1$ dimensions without emptying it, so that some products can be formed, but not enough to span the space of the higher order.

\begin{table*}[tb]
    \centering
    \includegraphics[width=\textwidth]{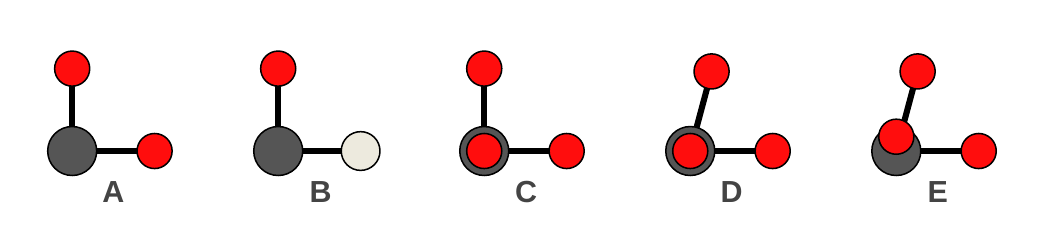}

    \begin{tabular}{c||c|c|c|c|c|c}
     & \multicolumn{6}{c}{Dimensionality} \\
    \textbf{Structure} & $\lambda=1$ & $\lambda=2$ & $\lambda=3$ & $\lambda=4$ & $\lambda=5$ & $\lambda=6$ \\ \hline
    \textbf{A} & 1 & 2 & 2 & 3 & 3 & 4 \\ \hline
    \textbf{B} & 2 & 3 & 4 & 5 & 6 & 7 \\ \hline
    \textbf{C} & 1 & 1 & 2 & 2 & 2 & 3 \\ \hline
    \textbf{D} & 2 & 3 & 4 & 5 & 6 & 7 \\ \hline
    \textbf{E} & 3 & 5 & 7 & 9 & 11 & 13 \\ \hline
    \end{tabular}
    \caption{Dimensionality of the space of each angular order $\lambda$ for the example structures shown above. Because of their symmetry, all structures except E have a reduced dimensionality, below $2\lambda + 1$, for every angular order. Although the dimensionality of each case can be derived analytically, here it is obtained numerically, by training a large MACE model ($L_{\max} = 4$, $\nu = 3$, four message-passing layers) to predict a basis of tensors of order $\lambda$, with a loss function that maximizes the rank of that basis.}
    \label{tab:si_basisincomp_dimensionalities}
\end{table*}

\begin{figure*}[tb]
    \centering
    \includegraphics[width=0.8\linewidth]{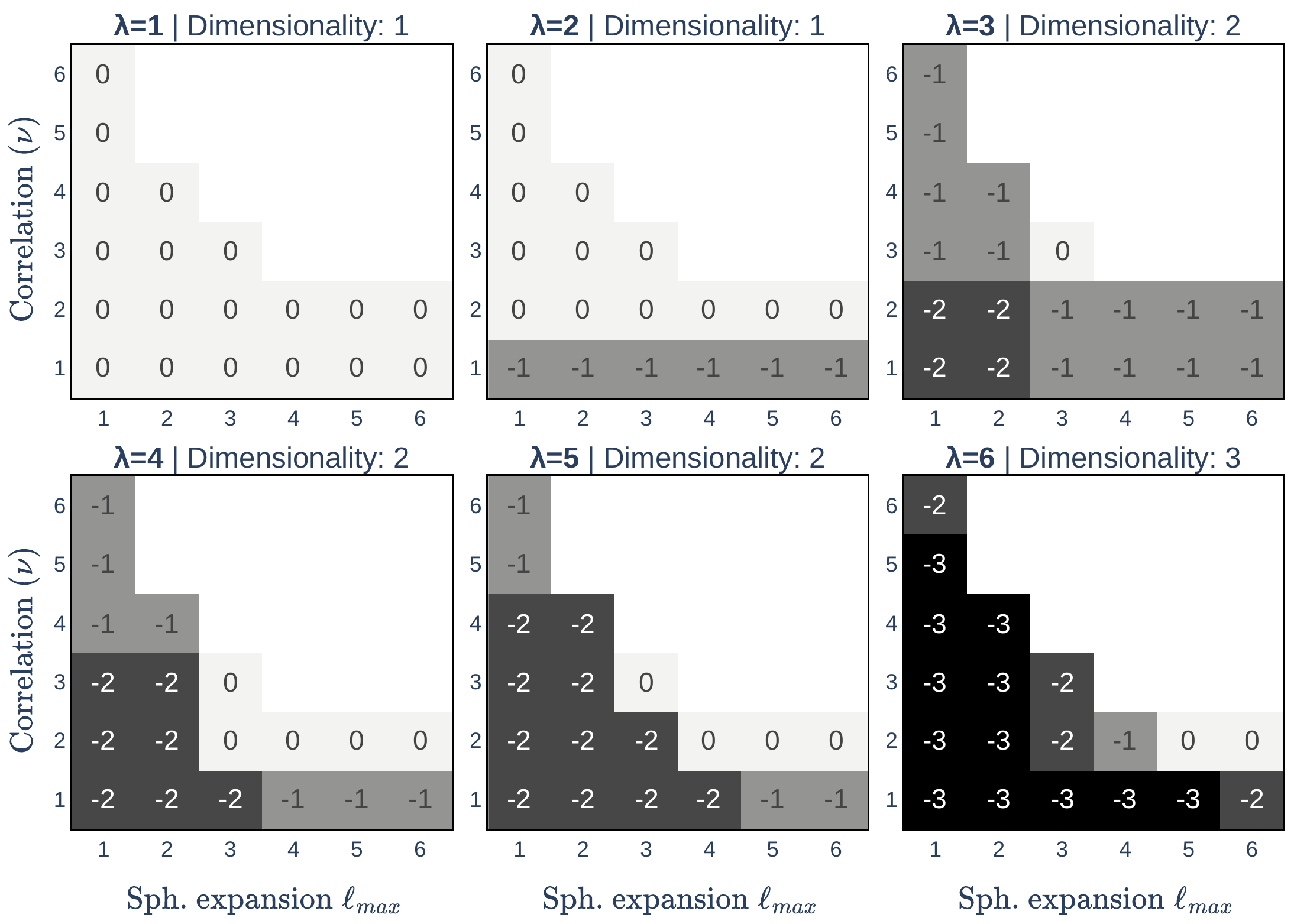}
    \caption{Coverage of the space of each angular order $\lambda$ for structure C of Table~\ref{tab:si_basisincomp_dimensionalities}, using only the spherical expansion and its products, as a function of the correlation order $\nu$ and of the spherical expansion order $\ell_{\max}$ ($L_{\max}$ in the text). The number in each cell is the number of missing dimensions for that $(\nu, \ell_{\max})$, zero meaning that the space, whose dimensionality is given above each panel, is fully covered. The results are obtained with the rank-maximizing loss of Table~\ref{tab:si_basisincomp_dimensionalities}, here with MACE models without message passing.}
    \label{fig:si_basis_incomp_struct_c_sweep}
\end{figure*}

Table~\ref{tab:si_basisincomp_dimensionalities} gives examples of structures in which symmetry restricts, without emptying, the spaces of each angular order, in a way that leaves the product basis incomplete.
The simplest case is structure A, whose $\lambda = 1$ space is one-dimensional, spanned by the bisector.
Products of $\lambda = 1$ features then yield a single independent $\lambda = 2$ tensor, whereas the $\lambda = 2$ space of this structure is two-dimensional, so that the product basis misses one dimension and cannot fit an arbitrary $\lambda = 2$ target.
The other structures can be analyzed in the same way.
Figure~\ref{fig:si_basis_incomp_struct_c_sweep} shows, for structure C, how many dimensions of each $\lambda$ space the product basis misses as a function of $L_{\max}$ and $\nu$: choosing them such that $L_{\max} \times \nu \geq \lambda$ does not guarantee a complete basis for $\lambda$, even when no angular channel vanishes completely.
For structures such as these, message passing removes the incompleteness by bringing in the missing dimensions from neighbors with less symmetric environments.
In periodic structures, however, the problem can persist despite message passing, for example in one-dimensional chains or honeycomb lattices.

\subsection{Training details of the FCC-Au benchmark}
\label{sec:si_fccau}

We take 100 structures from the dataset of Ref.~\onlinecite{qian2026equivariant}, which contains the on-site Hamiltonian blocks of 64-atom FCC-Au cells. Each model is trained on a single structure, with no validation set, and tested on the 99 others. The orbital basis carries 6, 5, 3 and 1 radial functions of $l = 0, \dots, 3$ per atom, so that the coupled on-site blocks span $\lambda = 0, \dots, 6$, with 71, 96, 81, 48, 20, 6 and 1 property channels respectively. Their magnitudes decay by about eight orders of magnitude from $\lambda = 0$ to $\lambda = 6$. After removing a per-type baseline from the $\lambda = 0$ block, every property channel of every block is therefore divided by its uncentered root-mean-square value on the training structure, so that all channels enter the mean-squared-error loss with equal weight; without this, neither model learns the $\lambda = 5$ and 6 blocks.

The MACE models carry 64 channels per irrep up to $l = 6$, three interaction layers, a cutoff of 5.0~\AA{} and eight Bessel radial functions, in double precision; two further models with $(L_{\max}, \nu) = (6, 1)$ carry 32 and 16 channels per irrep (Sec.~\ref{sec:si_fccau_timings}). They are trained for 2500 steps on the single structure, with Adam (AMSGrad), a learning rate of $10^{-2}$ reduced by a factor of 0.8 every 51 steps, a weight decay of $5\times10^{-7}$ and gradient clipping at norm 1.0. Every run had converged within 1250 steps.

The PET models use two GNN layers of two attention layers each, eight attention heads, a cutoff of 5.0~\AA{}, in single precision, with edge feature dimensions $d_{\text{PET}} = 40$, 64 and 96 and $d_{\text{node}} = d_{\text{ff}} = 2 d_{\text{PET}}$. At every step the training structure is replicated 128 times, each copy under an independent random rotation and inversion, and the loss is the mean squared error over the copies. They are trained for $5\times10^{4}$ steps with Adam, a linear warm-up over the first 1\% of steps to a learning rate of $10^{-3}$, held constant until 91\% of the steps and then decayed to zero with a cosine schedule, and gradient clipping at norm 1.0.

All models are trained once, with the same random seed, and evaluated with their final weights, on one NVIDIA H100 GPU per run: 0.5--0.7~h for MACE and 10--11~h for PET. The step budgets differ by design. MACE converges well within its budget, whereas PET, which must also learn the rotational orbit of the structure, had not saturated after $5\times10^{4}$ steps, so that its errors in Fig.~\ref{fig:angular_order_benchmark} are not converged values. Although the budget is far higher for PET than for MACE, the two learn different functions: PET must learn equivariance as well as the mapping from geometry to output. On real datasets with more diversity, the ratio of steps used to train PET compared to equivariant models is likely to be more balanced.

\subsection{Errors and inference cost of the FCC-Au models}
\label{sec:si_fccau_timings}

\begin{figure}[tb]
    \centering
    \includegraphics[width=\textwidth]{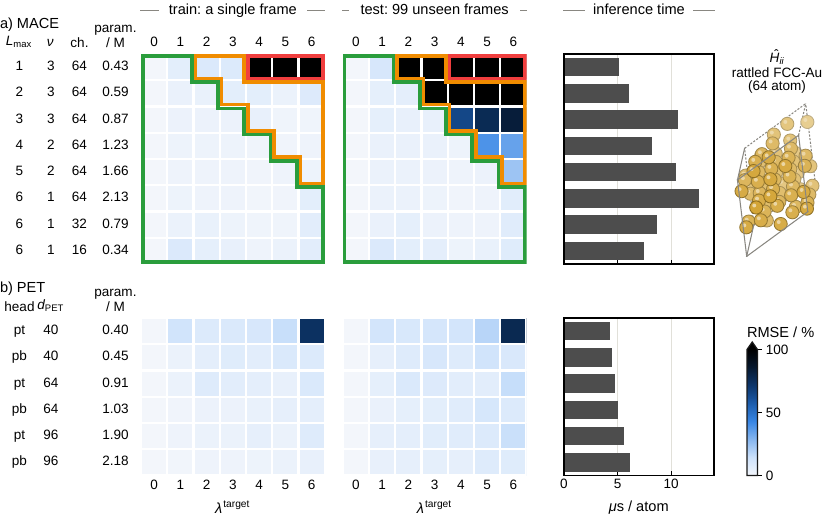}
    \caption{
    Errors and inference time of every model of the FCC-Au benchmark, one row per model: a) MACE, the six models of Fig.~\ref{fig:angular_order_benchmark} followed by $(L_{\max}, \nu) = (6, 1)$ with 32 and 16 channels per irrep (ch.), and b) PET. Left to right: the hyperparameters and parameter count of each model; its relative RMSE per target angular order $\lambda$ on the training structure and on the 99 unseen structures, with the colors and outlines of Fig.~\ref{fig:angular_order_benchmark}; and its inference time per atom (H100, single precision, 16\,384 atoms; MACE with cuEquivariance).
    }
    \label{fig:si_angular_order_timings}
\end{figure}

Figure~\ref{fig:si_angular_order_timings} collects, for every model of the benchmark, the per-$\lambda$ errors of Fig.~\ref{fig:angular_order_benchmark} and the inference time per atom. It adds two MACE models with $(L_{\max}, \nu) = (6, 1)$, the only expansion that represents every block directly, with the irrep width reduced from 64 to 32 and 16 channels (0.79 and 0.34~M parameters instead of 2.13~M). Both still fit and generalize every block: their test errors for $\lambda = 1, \dots, 6$ are 2.1--7.2\% and 2.5--9.1\%, against 1.2--7.0\% at 64 channels, and at every width the test error exceeds the training error by at most 1.3\%. The failure of the truncated models for $\lambda > L_{\max}$ is therefore a consequence of the angular resolution of their basis rather than of their parameter count, and an angularly complete model generalizes even at the size of the smallest PET model (0.40~M parameters). The price is computational. Their inference cost falls from 12.6 to 8.7 and 7.4~$\mu$s per atom, but stays above that of every PET model (4.3--6.2~$\mu$s per atom). With the reference e3nn implementation, the $L_{\max} = 6$ tensor products make even the 16-channel model an order of magnitude slower than the size-matched PET model (41 against 4.3~$\mu$s per atom), a gap that the fused cuEquivariance kernels reduce to a factor of 1.7.

Inference times are measured with the trained weights on one NVIDIA H100 GPU in single precision, without gradients, for a batch of 256 copies of the 64-atom training structure (16\,384 atoms), halved until it fits in memory for some of the e3nn MACE models, as the median of up to 30 iterations after 5 warm-up ones. MACE is timed with the fused cuEquivariance kernels, which are faster than its e3nn implementation for every model, by a factor of 2 to 12.

\section{Training protocol}
\label{sec:si_training}

\subsection{Backbone size and head/readout ablations}\label{sec:si_size_ablation}

For each dataset, models of increasing backbone size, labeled XS, S, M, and L, are trained following a simple heuristic, with hyperparameters given in Table~\ref{tab:backbone_hypers}.
The edge feature dimension $d_{\text{PET}}$ is increased in equal steps from 128 to 512, with the node and feed-forward dimensions, $d_{\text{node}}$ and $d_{\text{ff}}$, held at twice its value, and the number of transformer layers $n_{\text{TL}}$ raised from two to three for the largest model only.
The cutoff radius and the number of GNN layers are held fixed, at 5.0~\AA\ and two respectively, as all three datasets contain only molecular systems.

\begin{figure*}[tb]
    \centering
    \includegraphics[width=\linewidth]{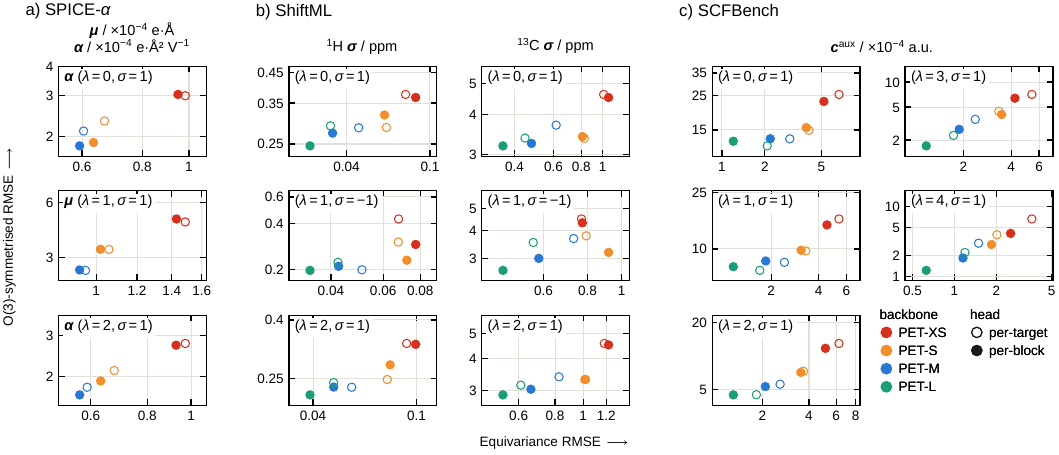}
    \caption{
    Per-irrep accuracy and equivariance of the backbone-size and head-type ablation summarized in Fig.~\ref{fig:equivariance_overview}a,b: $\mathrm{O}(3)$-symmetrized RMSE against equivariance RMSE for each irrep of a) SPICE-$\alpha$, b) the $^{1}$H and $^{13}$C shieldings of ShiftML, and c) SCFBench, for \texttt{PET-XS} to \texttt{PET-L} with per-target (hollow) and per-block (filled) heads. SCFBench errors are pooled over atomic types and radial functions, with atom type-conditioned readouts. Each panel has its own logarithmic axes.
    }
    \label{fig:si_size_ablation_irreps}
\end{figure*}

On each backbone both per-target and per-block heads are trained, with the number of parameters in the resulting head and readout modules reported in Table~\ref{tab:pet-head-readout-sizes}.
Per-block heads carry between 1.5 and 4.3 times as many parameters as per-target heads, increasing with the number of irrep blocks of the target, but remain small on the scale of the backbone: from around a third of its parameters for the XS density models to around a fifth for L.
The training hyperparameters and cost of each model are given in Table~\ref{tab:pet-training-hypers}.
The per-irrep accuracy and equivariance of every model, which underlie the summary of the main text, are shown in Fig.~\ref{fig:si_size_ablation_irreps}.

\begin{table}[tb]
\centering
\footnotesize
    \begin{tabular}{cccccc}
        \toprule
        \textbf{Model} & \textbf{Hyper.} & \textbf{XS} & \textbf{S} & \textbf{M} & \textbf{L} \\
        \midrule
        PET & $d_{\text{PET}}$ & 128 & 256 & 384 & 512 \\
         & $d_{\text{node}}$ & 256 & 512 & 768 & 1024 \\
         & $d_{\text{ff}}$ & 256 & 512 & 768 & 1024 \\
         & $n_{\text{TL}}$ & 2 & 2 & 2 & 3 \\
         & Param / M & 2.82 & 11.25 & 25.28 & 64.86 \\
        \bottomrule
    \end{tabular}
    \caption{
        Backbone architecture at different sizes. Parameter counts are for the backbone only, excluding the heads and readouts (Table~\ref{tab:pet-head-readout-sizes}). A fixed cutoff of 5.0~\AA{} is used for all in line with related works \cite{gonnheimer2026macemdp,kellner2025shiftml3,kellner2026shiftml4,liu2026scfbench}. As the datasets used span molecular systems only, we do not scale $n_{\text{GNN}}$ with model size as in Ref.~\cite{bigi2026pushing}, keeping it fixed at 2 layers.
    }
    \label{tab:backbone_hypers}
\end{table}

\begin{table}[tb]
  \centering
  \caption{Number of learnable parameters (in millions) in the PET head+readout modules for each dataset, model size and head type (per-target, per-block). Values are for the standard readout for SPICE-$\alpha$ and ShiftML, and for the species-conditioned readout for SCFBench.}
  \label{tab:pet-head-readout-sizes}
  \begin{tabular}{lcc}
    \toprule
    Size & Per-target & Per-block \\
    \midrule
    \multicolumn{3}{l}{\textit{SPICE-$\alpha$}} \\
    XS & 0.33 & 0.50 \\
    S & 1.32 & 1.98 \\
    M & 2.96 & 4.44 \\
    \midrule
    \multicolumn{3}{l}{\textit{ShiftML}} \\
    XS & 0.17 & 0.50 \\
    S & 0.66 & 1.98 \\
    M & 1.49 & 4.44 \\
    L & 2.64 & 7.89 \\
    \midrule
    \multicolumn{3}{l}{\textit{SCFBench -- def2-universal-jfit}} \\
    XS & 0.30 & 0.96 \\
    S & 0.93 & 3.56 \\
    M & 1.89 & 7.80 \\
    L & 3.17 & 13.67 \\
    \midrule
    \multicolumn{3}{l}{\textit{SCFBench -- etb-2.0}} \\
    L & -- & 14.90 \\
    \bottomrule
  \end{tabular}
\end{table}

\begin{table}[tb]
  \centering
  \small
  \setlength{\tabcolsep}{3.5pt}
  \caption{PET training hyperparameters and cost for each dataset and model size. All models were trained on GH200 nodes (4 GPUs per node) with the Adam optimizer, an MSE loss on scaled targets, a linear learning-rate warm-up over the first 1\% of optimizer steps followed by cosine decay to zero, gradient-norm clipping at 1.0, no weight decay, single precision and a fixed random seed. The batch size is the effective batch over all GPUs, with 256 structures per GPU (170 for SPICE-$\alpha$ M, 128 for ShiftML L). The selected model is the checkpoint with the lowest product over targets of the validation MAE. Wall time is the average over the head/readout ablations at that backbone size, for generating the ablation results in Fig.~\ref{fig:equivariance_overview}a.}
  \label{tab:pet-training-hypers}
  \begin{tabular}{llrrrrr}
    \toprule
    \multicolumn{1}{l}{Dataset} & Size & Epochs & GPUs & \begin{tabular}[b]{@{}c@{}}Batch\\size\end{tabular} & \begin{tabular}[b]{@{}c@{}}Peak\\LR\end{tabular} & \begin{tabular}[b]{@{}c@{}}Wall\\time (h)\end{tabular} \\
    \midrule
    SPICE-$\alpha$ & XS & 500 & 32 & 8192 & $5\times 10^{-4}$ & 14.6 \\
                   & S & 500 & 32 & 8192 & $5\times 10^{-4}$ & 21.4 \\
                   & M & 500 & 48 & 8160 & $5\times 10^{-4}$ & 23.4 \\
    \midrule
    ShiftML & XS & 2000 & 16 & 4096 & $3\times 10^{-4}$ & 3.3 \\
            & S & 2000 & 16 & 4096 & $3\times 10^{-4}$ & 4.3 \\
            & M & 2000 & 16 & 4096 & $3\times 10^{-4}$ & 6.0 \\
            & L & 2000 & 32 & 4096 & $3\times 10^{-4}$ & 5.3 \\
    \midrule
    SCFBench & XS & 5000 & 16 & 4096 & $2\times 10^{-4}$ & 4.9 \\
             & S & 5000 & 16 & 4096 & $2\times 10^{-4}$ & 5.6 \\
             & M & 5000 & 16 & 4096 & $2\times 10^{-4}$ & 6.5 \\
             & L & 5000 & 16 & 4096 & $2\times 10^{-4}$ & 8.9 \\
    \bottomrule
  \end{tabular}
\end{table}

\subsection{Geometry embeddings}
\label{sec:si_geometry_embeddings}

In Ref.~\onlinecite{domina2026unconstrained} we found that standard PET fails to learn an isolated target of high angular order, and that this is remedied by replacing its geometry embeddings with a solid spherical harmonics (SSH) expansion of the edge vectors, which supplies high-$\lambda$ content from the start of training.
The targets of this work are instead defined on a hierarchy of angular orders, from $\lambda = 0$ up to $\lambda_{\text{max}}^{\text{target}}$.
For these, SSH embeddings can bring forward the onset of learning of the high-$\lambda$ terms, but with either embedding the higher orders are learned sequentially, as their gradients accumulate gradually, and the converged models perform similarly.
We therefore use the standard embeddings throughout.

\subsection{Baselines and scales}\label{sec:si_scales}

For each dataset, atomic baselines and scales of the training data are computed.
For a generic equivariant target on the coupled angular momentum basis, $y^{\lambda \sigma (Z)}_{A (i) \mu q}$, the atomic baseline is computed as the atomic-type-resolved mean over the dataset of the invariant $(\lambda = 0, \sigma = +1)$ irreps.

For targets with an atomic decomposition (chemical shieldings and electron density coefficients), this is computed as a simple mean over atomic samples for each atom type $Z$ and property $q$:
\begin{equation}
    b^{0 1 Z}_{q} = \frac{ \sum_{A i} y^{0 1 Z}_{A i 0 q} \delta_{Z_i Z} }{ \sum_{A i} \delta_{Z_i Z} } \ ,
\end{equation}
These are then subtracted from the invariant terms of the target as:
\begin{equation}
    y^{0 1 Z}_{A i 0 q} \rightarrow y^{0 1 Z}_{A i 0 q}  - b^{0 1 Z}_{q}
\end{equation}

For global (per-system) quantities, the following linear problem is solved to estimate the atomic contribution to the global invariant quantity, with $N_{AZ}$ the number of atoms of type $Z$ in system $A$:
\begin{equation}
    \{b^{0 1 Z}_{q}\} = \arg\min_{\{b^{Z}\}} \sum_{A} \left( y^{0 1}_{A 0 q} - \sum_{Z} N_{AZ} \, b^{Z} \right)^2 ,
\end{equation}
with the baseline removal as follows:
\begin{equation}
    y^{0 1}_{A 0 q} \rightarrow y^{0 1}_{A 0 q} - \textstyle\sum_{Z} N_{AZ} \, b^{0 1 Z}_{q}
\end{equation}

Scales are then computed on the baselined targets over the training set, as uncentered (and therefore equivariance-preserving) standard deviations. First, a scale is computed that aggregates the reference data of all irreps: a single scale for per-system targets, and a per-species scale for atomic targets.

For global targets, each structure's target is first divided by its number of atoms $N_A$, and this is a single scalar:
\begin{equation}
    s = \sqrt{ \frac{ \sum_{\lambda \sigma A \mu q} \left( y^{\lambda \sigma}_{A \mu q} / N_A \right) ^ 2 } { \sum_{\lambda \sigma A \mu q} 1 } }
\end{equation}
and for atomic targets this is one scalar per atomic type:
\begin{equation}
    s^{Z} = \sqrt{ \frac{ \sum_{\lambda \sigma A i \mu q} \left( y^{\lambda \sigma Z}_{A i \mu q} \delta_{Z_i Z} \right) ^ 2 } { \sum_{\lambda \sigma A i \mu q} \delta_{Z_i Z} } }
\end{equation}

At this point, large invariant baselines have been removed and the global scales of the targets normalized. In a multi-target training regime, e.g., when jointly learning dipoles and polarizabilities, this removes the units from the targets, ensuring that one does not dominate the loss purely due to different physical scales, thus removing the need for hand-tuning their weighting in the loss.

For generic equivariant targets, however, different irreps and basis functions can still have vastly different relative scales: in general the magnitude of spherical components typically follows a decay as the angular order increases, and different nuclei naturally have very different scales in chemical shielding tensors or atomic basis functions. Typically in ML, normalizing all targets to unit standard deviation ensures stable dynamics during training, though the relative scales of different properties and irreps should still enter the loss to ensure the model properly weights these terms by their physical importance.

We therefore compute a second set of scales, conditioned by sparse block (O(3)-irrep and atom type, $(\lambda, \sigma, Z)$) and property ($q$), that provide a non-learnable unscaling of the model predictions prior to loss computation. They are computed on the targets already divided by the global scale(s), so that they are relative factors and the full scale of an entry is $s \, s^{\lambda \sigma}_{q}$ (or $s^{Z} s^{\lambda \sigma Z}_{q}$). For global targets, these are computed as:
\begin{equation}
    s^{\lambda \sigma}_{q} = \sqrt{ \frac{ \sum_{A \mu} \left( y^{\lambda \sigma}_{A \mu q} / (N_A \, s) \right) ^ 2 } { \sum_{A \mu} 1 } }
\end{equation}
and for atomic targets as:
\begin{equation}
    s^{\lambda \sigma Z}_{q} = \sqrt{ \frac{ \sum_{A i \mu} \left( y^{\lambda \sigma Z}_{A i \mu q} \delta_{Z_i Z} / s^{Z} \right) ^ 2 } { \sum_{A i \mu} \delta_{Z_i Z} } }
\end{equation}
For the electron density, the property index $q$ runs over the radial functions $n$ of each atomic type, so that every radial channel of every irrep has its own scale.

\subsection{Effect of the scaling convention}

\begin{figure}[tb]
    \centering
    \includegraphics[width=\linewidth]{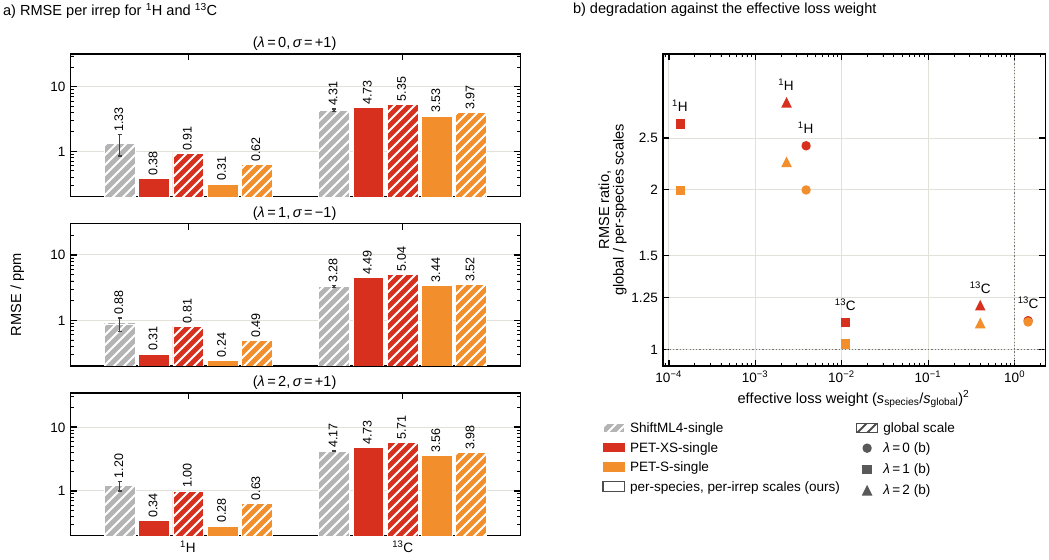}
    \caption{
    Effect of the target-scaling convention of ShiftML4 on PET. a) Per-atom RMSE of each irrep of the $^{1}$H and $^{13}$C shielding tensors for \texttt{PET-XS} and \texttt{PET-S} single models trained with our per-species, per-irrep scales (solid) and with the single global scale of ShiftML4 (hatched), and for the single ShiftML4 models~\cite{kellner2026shiftml4} (mean $\pm$ standard deviation over the members of its ensemble), which use the global scale. ShiftML4 is evaluated on the 990-structure subset of the test set used in the main text and PET on the full test set, on which its errors differ by 1--4\%. b) Ratio of the RMSE with the global scale to that with per-species scales, against the effective loss weight $(s_{\text{species}} / s_{\text{global}})^{2}$ of each atomic type and irrep.
    }
    \label{fig:si_sml4_baseline_ablation}
\end{figure}

The per-block scales matter most when the targets of different atomic types differ in scale by orders of magnitude, as for chemical shieldings: the $(0, +1)$ scale of $^{1}$H is 4.8~ppm, against 92~ppm for $^{13}$C.
The ShiftML4 models~\cite{kellner2026shiftml4} are instead trained with a single global scale, pooled over all atomic types and irreps (76.8~ppm).
In a loss on scaled targets, a block whose own scale $s_{\text{block}}$ lies below the global one $s_{\text{global}}$ is effectively de-weighted by $(s_{\text{block}} / s_{\text{global}})^{2}$, a factor of $10^{-4}$ to $10^{-3}$ for the irreps of $^{1}$H.
To isolate this effect, we retrained \texttt{PET-XS} and \texttt{PET-S} exactly as in the main text, but with the atomic baseline and target scaler of the published ShiftML4 model (Fig.~\ref{fig:si_sml4_baseline_ablation}).
The baselines of the two agree to around 0.01~ppm for $^{1}$H and $^{13}$C, so that the change is effectively one of scaling.
With the global scale, the $^{1}$H errors of PET rise by a factor of two (\texttt{PET-S}) to three (\texttt{PET-XS}) in every irrep, while those of $^{13}$C, whose scale is close to the global one, rise by around 20\% at most.
Even so, the single ShiftML4 models remain no more accurate than \texttt{PET-XS} trained with the global scale for $^{1}$H, while for $^{13}$C they fall between \texttt{PET-XS} and \texttt{PET-S} trained with the global scale in the $(0, +1)$ and $(2, +1)$ blocks.
The scaling convention thus explains part, but not all, of the larger $^{1}$H errors of ShiftML4 reported in the main text.

\subsection{Loss functions}\label{sec:si_losses}

For quantities learned \textit{directly}, an MSE loss, averaged over all components, is computed between the baselined and globally scaled targets and the irrep- and property-unscaled predictions:
\begin{align*}
    \text{Targets:} \ y^{\lambda \sigma (Z)}_{A (i) \mu q} \rightarrow \frac{y^{\lambda \sigma (Z)}_{A (i) \mu q} - b^{01(Z)}}{s^{(Z)}} \\
    \text{Predictions:} \ \tilde{y}^{\lambda \sigma (Z)}_{A (i) \mu q} \rightarrow s^{\lambda \sigma (Z)}_{q} \tilde{y}^{\lambda \sigma (Z)}_{A (i) \mu q} \\
\end{align*}

The loss is averaged over the systems $A$ in the minibatch and, for atomic quantities, over the atoms $i$ of each system. For global targets, both targets and predictions are further divided by $N_A$.

For learning a scalar field \textit{indirectly} by targeting its basis-set expansion coefficients, we use an L2 loss that minimizes the error relative to the reference scalar field. This is computed independently for each system as:
\begin{equation}
    \mathcal{L}_A = \int d^3 \textbf{r} \left|  \rho^{\text{ML}}_A (\textbf{r}) - \rho^{\text{ref}}_A (\textbf{r}) \right| ^ 2
\end{equation}
with a simple mean reduction over systems in the batch. Choosing the reference scalar field as the one reconstructed from the reference coefficients, the loss can be reformulated on the auxiliary basis, with real-space integrals implicitly contained within the overlap metric:
\begin{equation} \label{eq:metric_loss_dcSdc}
    \mathcal{L}_A = \left( \Delta \textbf{c}^{\text{aux}}_A \right)^T \cdot \hat{S}_A \cdot \Delta \textbf{c}^{\text{aux}}_A \ , 
\end{equation}
where $\textbf{c}^{\text{aux}}_A$ is the set of all coefficients $\{c^{\lambda \sigma}_{i \mu n}\}$ for system $A$ and $\Delta \textbf{c}^{\text{aux}}_A = \tilde{\textbf{c}}^{\text{aux}}_A - \textbf{c}^{\text{aux}}_A$ is the difference between the predicted and reference coefficients. As $\hat{S}_A$ depends only on nuclear geometry (and the auxiliary basis definition, which can be considered a model hyperparameter) but not on the model weights, the metrics can be pre-computed prior to training and loaded, or dynamically computed and cached during training.
This is the S density loss of the main text, with $\textbf{M}_A = \hat{S}_A$. Replacing the overlap with the Coulomb metric of the auxiliary basis, $\hat{J}_{A, kl} = \iint d\textbf{r} \, d\textbf{r}' \, \phi_k(\textbf{r}) \phi_l(\textbf{r}') / |\textbf{r} - \textbf{r}'|$, gives the J density loss, which measures the error in the electrostatic self-repulsion of the density difference.

\subsection{Data augmentation}

PET is not constrained to exact O(3)-equivariance, so these symmetries must be learned \textit{via} data augmentation. During training, random rotations and inversions are applied separately to each system in the batch and to its targets. In the usual case, a forward pass is evaluated on the transformed system and the loss evaluated between the prediction and the transformed target.

Where the metric loss in Eq.~\eqref{eq:metric_loss_dcSdc} is used, the model is evaluated on transformed systems but the targets remain untransformed, with the loss computed between them and the predicted $\tilde{\textbf{c}}^{\text{aux}}_A$ back-transformed to the original frame. This avoids applying expensive augmentation on $\{\hat{S}_A\}$ during training.

\subsection{Electron-density models}

\subsubsection{Density-loss fine-tuning.}\label{sec:si_density_ft}

Each density-loss stage starts from the previous checkpoint with every parameter trainable, and uses Adam, gradient clipping at norm 1.0 and a cosine decay of the learning rate to zero. Starting from the models of Table~\ref{tab:pet-training-hypers}, trained for 5000 epochs on the coefficients, the S- and J-loss fine-tuning of the main text runs for 1000 epochs at a peak learning rate of $2\times10^{-4}$, without warm-up, with a global batch of 4096 structures on 16 GH200 GPUs. The extended \texttt{etb-2.0} model continues for a further 5000 fine-tuning epochs with the J-loss in 3 stages: 1000 epochs at a LR of $10^{-4}$, 2000 epochs at $5\times10^{-5}$ and 2000 epochs at $2.5\times10^{-5}$. The metric matrices are computed dynamically during training with PySCF~\cite{sun2020pyscf}.

\subsubsection{Reference coefficients.}\label{sec:si_ref_coeffs}

For a reference density matrix $\textbf{D}$ in the def2-SVP orbital basis, the S-fitted coefficients distributed with SCFBench are $\textbf{c}^{\text{aux}, S} = \hat{S}^{-1} \textbf{w}^{S}$, with $w^{S}_P = \sum_{ij} D_{ij} \, (ij P)$ the three-center overlap integrals. The J-fitted coefficients are $\textbf{c}^{\text{aux}, J} = \hat{J}^{-1} \textbf{w}^{J}$, with $w^{J}_P = \sum_{ij} D_{ij} \, (ij|P)$ the three-center Coulomb integrals, computed with PySCF in double precision. No electron-number constraint is imposed in the fits performed here. The J-fitted \texttt{def2-universal-jfit} training set, which was generated separately, instead corresponds to the electron-number-constrained Coulomb fit of Sec.~\ref{sec:si_electron_correction}, $\textbf{c}^{\text{aux}, J} = \hat{J}^{-1} \textbf{w}^{J} + \kappa \hat{J}^{-1} \textbf{q}$, where $q_P$ is the integral of auxiliary function $P$ and $\kappa$ is fixed so that the density integrates to the electron count $\mathrm{tr}(\textbf{D}\textbf{S})$ of the reference density matrix, with $\hat{J}$ regularized by adding $10^{-8}$ to its diagonal and inverted on the eigenvectors with eigenvalues above $10^{-10}$. Its out-of-distribution test set, refitted here, is unconstrained. The S-fitted densities are therefore not neutral, with a mean absolute charge of 0.083 and 0.020~$e$ for \texttt{def2-universal-jfit} and \texttt{etb-2.0} on the in-distribution test structures, whereas the J-fitted densities, constrained or not, are neutral to within $6\times10^{-4}$~$e$, as the Coulomb metric strongly penalizes the long-range potential of any net charge of the residual.

\subsubsection{SCF evaluation.}

All one-shot and SCF calculations use PySCF on CPU in double precision, with exact two-electron integrals and no renormalization of the predicted electron count. Both SCF runs, from the predicted guess and from \texttt{minao}, are run with the same code and the settings with which the SCFBench references were generated (convergence threshold $10^{-10}$~Ha, at most 50 cycles, a (99, 590) atomic grid). The evaluation code released with SCFBench instead uses \texttt{gpu4pyscf} with the default PySCF threshold and grid, and evaluates a random 1\% of the in-distribution and all of the out-of-distribution test set. Here every model is evaluated on the same 250 structures per test set, both SCF runs of each structure are checked to converge to the same energy within $10^{-6}$~Ha (all do), and NequIP-L is run with the model and post-processing code of SCFBench before passing through the same evaluation. On the full test sets, and with the \texttt{minao} cycle counts distributed with SCFBench as the baseline, the relative SCF cycles of the S-fitted reference coefficients lie 2--3\% above those of Ref.~\onlinecite{liu2026scfbench}.

\section{Improving equivariance errors}
\label{sec:si_improving_equi}

Figure~\ref{fig:si_ensemble_equivariance} extends the ensemble analysis of the main text to $^{1}$H and to every irrep of both nuclei.

\subsection{Last-layer feature character}
\label{sec:si_llf_character}

The $\mathrm{O}(3)$ character of the LLF in Fig.~\ref{fig:equivariance_overview}c is the normalized character power of Ref.~\onlinecite{domina2026unconstrained}, computed on the features that enter the affine readout of each head: 1024 node channels and 512 edge channels, pooled over neighbors with the cutoff weights. The per-block model has one head for each irrep of the shielding tensor, whereas the per-target model shares one head among all three. For central species $Z$, head $h$ and character $(L, \sigma)$, the powers of the node and edge features are summed over all atoms of that species before normalizing,
\begin{equation}
    B^{L\sigma}_{Zh} =
    \frac{\sum_{i:Z_i=Z}
        \left(
        \left\lVert\Pi_{L\sigma}\boldsymbol{\phi}^{\mathrm{node}}_{ih}\right\rVert^2
        +
        \left\lVert\Pi_{L\sigma}\boldsymbol{\phi}^{\mathrm{edge}}_{ih}\right\rVert^2
        \right)}
    {\sum_{i:Z_i=Z}
        \left\langle
        \left\lVert\boldsymbol{\phi}^{\mathrm{node}}_{ih}(g)\right\rVert^2
        +
        \left\lVert\boldsymbol{\phi}^{\mathrm{edge}}_{ih}(g)\right\rVert^2
        \right\rangle_g},
    \label{eq:si_llf_character_fraction}
\end{equation}
where $\Pi_{L\sigma}$ projects onto the character $(L, \sigma)$ and the average runs over the operations $g$ of the grid, excluding the bias columns. The projections are computed on the grid of degree 14 (Sec.~\ref{sec:si_quadrature}), which resolves $L = 0, \dots, 7$ for both parities, of which $L \le 4$ are shown. They describe the native features, not the prediction error.

\subsection{Linear purification}
\label{sec:si_purification}

Following Ref.~\onlinecite{domina2026unconstrained}, we refit only the final affine readout, keeping the backbone, the heads, the target scales and the baselines fixed. For species $Z$ and target irrep $\alpha = (\lambda, \sigma)$, let $\boldsymbol{x}_{ir}$ collect the node and edge features of atom $i$ in the structure transformed by the operation $g_r$, with a bias column for each. The prediction transformed back to the original frame, and its average over the grid, are
\begin{equation}
    \boldsymbol{z}_{ir}(\boldsymbol{W}) =
    \rho_\alpha^{-1}(g_r)
    \left[
        s_{Z\alpha}\boldsymbol{W}_{Z\alpha}\boldsymbol{x}_{ir}
        + \boldsymbol{b}_{i\alpha}(g_r)
    \right],
    \qquad
    \overline{\boldsymbol{z}}_i(\boldsymbol{W}) =
    \sum_r \omega_r\boldsymbol{z}_{ir}(\boldsymbol{W}),
    \label{eq:si_purification_prediction}
\end{equation}
with $s_{Z\alpha}$ the target scale, $\boldsymbol{b}_{i\alpha}$ the baseline, $\rho_\alpha$ the representation of the irrep, and $\omega_r$ the quadrature weights, which sum to one. For each model, species and irrep, the $2\lambda+1$ components are fitted jointly by minimizing
\begin{equation}
    \mathcal{J}_{Z\alpha}(\boldsymbol{W};\gamma,\varepsilon)
    = \frac{1}{N_{Z\alpha}}
    \sum_{i=1}^{N_{Z\alpha}} \sum_r\omega_r
    \left[
        \left\lVert\boldsymbol{z}_{ir}(\boldsymbol{W})-\boldsymbol{y}_i\right\rVert^2
        +\gamma
        \left\lVert\boldsymbol{z}_{ir}(\boldsymbol{W})-\overline{\boldsymbol{z}}_i(\boldsymbol{W})\right\rVert^2
    \right]
    +\varepsilon\left\lVert\boldsymbol{W}_{Z\alpha}\right\rVert_F^2 ,
    \label{eq:si_purification_objective}
\end{equation}
the squared error over the grid, with a penalty of weight $\gamma$ on the equivariance error and a ridge term of weight $\varepsilon$, over the $N_{Z\alpha}$ atoms of species $Z$. Whereas the native readout is shared across species, the refit gives each of the eight species present in the validation set its own matrix, with a shared one for Na, Mg and Ca. As $\mathcal{J}_{Z\alpha}$ is quadratic in $\boldsymbol{W}$, the moments of the transformed features are accumulated once, in double precision, and each choice of $(\gamma, \varepsilon)$ only requires solving the regularized normal equations.

The moments are computed on all 50\,114 training structures with the grid of degree 14, and $\gamma$ and $\varepsilon$ are selected on a disjoint validation set of 1834 structures. $\gamma$ takes 242 values, zero and logarithmically spaced values from $10^{-4}$ to $10^{4}$, and for each $\gamma$ the ridge $\varepsilon$ is chosen from a logarithmic grid relative to the median eigenvalue of the normal matrix, both by the lowest validation total RMSE. The refit with $\gamma = 0$ is therefore a ridge refit of the readout, not the native one. The purified readouts are evaluated on the 990 test structures of Sec.~\ref{sec:si_nmr} with the grid of degree 7, as for the native models.

For H and C (Table~\ref{tab:si_purification_hc}), purification lowers the equivariance RMSE by 1.9 to 10.7\% for both heads and every irrep, while the total RMSE changes by less than 1\%. These are paired changes of a fixed model, and are smaller than the variation of the equivariance error of \texttt{PET-L} between training seeds (7 to 13\%, Sec.~\ref{sec:improving_equi}). A comparison with the grid of degree 9 on one structure changed a few N and O values by more than 1\%, so that these values are estimates on a finite grid.

\begin{table*}[t]
\centering
\small
\setlength{\tabcolsep}{4pt}
\caption{Equivariance RMSE $A$ and total RMSE $R$, in ppm, of the native and purified readouts of the per-block (PB) and per-target (PT) \texttt{PET-L} models for H and C, on the 990 test structures with the grid of degree 7. $\Delta=100(\mathrm{purified}/\mathrm{native}-1)$, with paired 95\% bootstrap intervals from 2000 resamples of whole test structures, conditional on the selected readouts and the grid.}
\label{tab:si_purification_hc}
\begin{tabular}{lllrrrr}
\hline\hline
Head & Species & $(\lambda,\sigma)$ & $A$: native $\to$ purified & $\Delta A$ [\%] & $R$: native $\to$ purified & $\Delta R$ [\%] \\
\hline
PB & H & $(0,+1)$ & $0.0268 \to 0.0259$ & $-3.06\;[-3.16,-2.96]$ & $0.2476 \to 0.2470$ & $-0.23\;[-0.37,-0.08]$ \\
PB & H & $(1,-1)$ & $0.0340 \to 0.0309$ & $-9.11\;[-9.23,-9.00]$ & $0.1994 \to 0.1986$ & $-0.40\;[-0.67,-0.14]$ \\
PB & H & $(2,+1)$ & $0.0389 \to 0.0376$ & $-3.46\;[-3.56,-3.37]$ & $0.2219 \to 0.2217$ & $-0.09\;[-0.18,+0.01]$ \\
PB & C & $(0,+1)$ & $0.3546 \to 0.3458$ & $-2.48\;[-2.55,-2.41]$ & $3.2107 \to 3.2078$ & $-0.09\;[-0.24,+0.07]$ \\
PB & C & $(1,-1)$ & $0.4604 \to 0.4334$ & $-5.87\;[-5.98,-5.76]$ & $2.6957 \to 2.6852$ & $-0.39\;[-0.53,-0.25]$ \\
PB & C & $(2,+1)$ & $0.5298 \to 0.5092$ & $-3.90\;[-3.98,-3.82]$ & $2.9267 \to 2.9152$ & $-0.40\;[-0.48,-0.31]$ \\
\hline
PT & H & $(0,+1)$ & $0.0335 \to 0.0328$ & $-2.18\;[-2.27,-2.09]$ & $0.2921 \to 0.2928$ & $+0.22\;[-0.03,+0.42]$ \\
PT & H & $(1,-1)$ & $0.0422 \to 0.0404$ & $-4.41\;[-4.49,-4.34]$ & $0.2259 \to 0.2255$ & $-0.16\;[-0.26,-0.06]$ \\
PT & H & $(2,+1)$ & $0.0480 \to 0.0451$ & $-6.02\;[-6.18,-5.87]$ & $0.2459 \to 0.2451$ & $-0.32\;[-0.40,-0.26]$ \\
PT & C & $(0,+1)$ & $0.4463 \to 0.4379$ & $-1.86\;[-1.94,-1.79]$ & $3.4054 \to 3.3999$ & $-0.16\;[-0.42,+0.15]$ \\
PT & C & $(1,-1)$ & $0.5612 \to 0.5009$ & $-10.74\;[-10.94,-10.52]$ & $3.5817 \to 3.5471$ & $-0.97\;[-1.15,-0.76]$ \\
PT & C & $(2,+1)$ & $0.6101 \to 0.5637$ & $-7.61\;[-7.75,-7.47]$ & $3.1991 \to 3.1747$ & $-0.76\;[-0.88,-0.65]$ \\
\hline\hline
\end{tabular}
\end{table*}

\subsection{Orientation grids}
\label{sec:si_quadrature}

Where an exactly symmetric quantity or an equivariance error is computed on an $\mathrm{O}(3)$ quadrature, we use the grids of \texttt{metatomic.torch.o3}: a Lebedev grid of the rotation axis combined with equispaced in-plane rotations and with inversion, with quadrature weights. For the spectra, the grid of integration degree 9 (760 operations) is converged: for the most orientation-sensitive Raman spectrum, the per-atom equivariance error of $\partial\boldsymbol{\alpha}_i/\partial Q_k$ changes by at most 0.6\% between degree 9 and 13, against 7\% between degree 5 and 9, while 24 random orientations deviate from degree 13 by up to 29\%.
The grid of degree $L$ combines the smallest Lebedev rule of order at least $L$ with $L+1$ in-plane rotations and with inversion, and integrates Wigner-D matrices exactly up to degree $L$. The equivariance and $\mathrm{O}(3)$-symmetrized errors of Fig.~\ref{fig:equivariance_overview} are computed on the grid of degree 7 (26 Lebedev nodes $\times$ 8 in-plane rotations $\times$ 2, i.e.\ 416 operations) for SPICE-$\alpha$ and ShiftML and of degree 9 (760 operations) for SCFBench, on 1000 test structures drawn at random from each test set, of which 990 remain for ShiftML-molecules after the outlier filter of Sec.~\ref{sec:si_nmr}. The ensembles of Fig.~\ref{fig:equivariance_overview}d, and the purified readouts of Fig.~\ref{fig:equivariance_overview}c, are evaluated on the same 416-operation grid. In Fig.~\ref{fig:isotropics}, the parity plots of panel a use the predictions in the original orientation, while the distributions of panels b and c, which concern only the invariant $\sigma_{\text{iso}}$, are computed on all test structures with a coarser set of 76 operations: the 38 nodes of the order-9 Lebedev rule, a single in-plane angle, and inversion. With a single in-plane angle per axis, this set samples only a two-dimensional slice of $\mathrm{SO}(3)$ and is not an exact quadrature. For $\sigma_{\text{iso}}$ it nonetheless reproduces, to within 3\% for every model and for both $^{1}$H and $^{13}$C, the per-atom invariance errors that the 416-operation quadrature gives on the 990-structure subset.

\begin{figure}
    \centering
    \includegraphics[width=\linewidth]{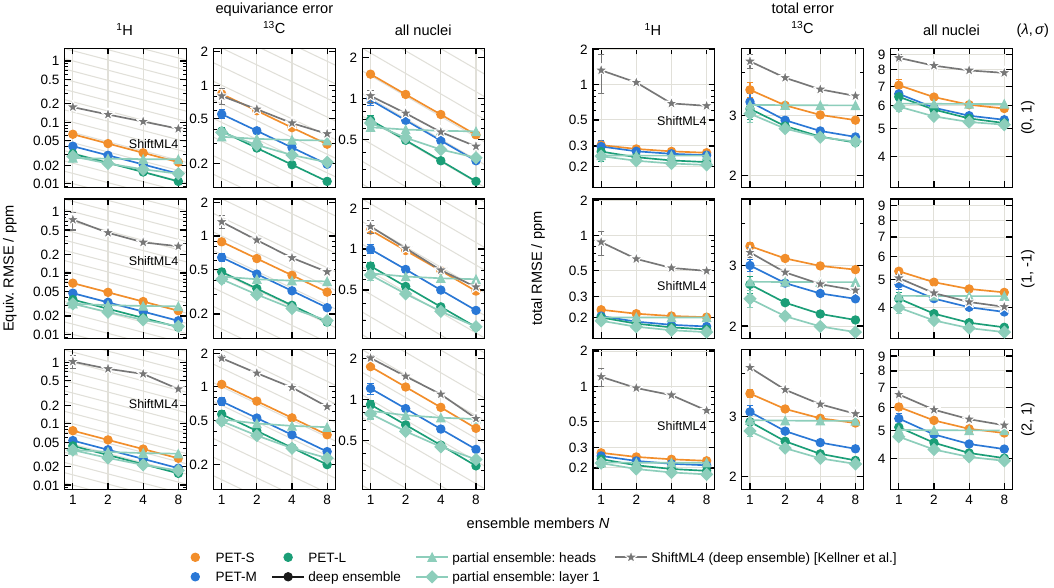}
    \caption{
    Equivariance error (left) and total error (right) of the shielding tensor of $^{1}$H, $^{13}$C and all nuclei, for each target irrep $(\lambda, \sigma)$, against the number $N$ of ensemble members, on the 990-structure subset of the ShiftML-molecules test set. The all-nuclei columns give the RMSE over every atom of these structures (H, C, N, O, F, P, S and Cl). Deep ensembles of \texttt{PET-S}, \texttt{PET-M} and \texttt{PET-L} are compared with partial \texttt{PET-L} ensembles branching at the heads or after the first GNN layer, and with the 8-member ShiftML4 ensemble~\cite{kellner2026shiftml4} and its sub-ensembles. For $N < 8$, errors are averaged over disjoint subsets of the members; error bars at $N = 1$ give the standard deviation over individual models. The equivariance panels carry $N^{-1/2}$ guide lines. All errors are RMSEs in ppm in the spherical basis.
    }
    \label{fig:si_ensemble_equivariance}
\end{figure}

\subsection{Ensemble training details}
\label{sec:si_training_ensembles}

Member 0 of each deep ensemble is the single per-block model of the main text for that backbone size. Members 1--7 are trained from scratch with identical settings (Table~\ref{tab:pet-training-hypers}) except the random seed, which sets the weight initialization, the order of the training data, and the random rotations and inversions used for augmentation. The ensemble prediction is the mean over members of every spherical component, formed before any error is computed.

Both partial ensembles branch from the trained single \texttt{PET-L} model, which is also their member 0. For the variant branched after the first GNN layer, used in the main text, each further member copies that model and re-initializes, with its own seed, the second GNN layer, the message-combination layers, the heads and the readouts, keeping the element embeddings, the first GNN layer, the baseline and the scaler. Only the re-initialized modules are then trained, separately for each member, with the mean-squared-error loss, for 400 epochs at a peak learning rate of $3\times10^{-4}$ with warm-up and cosine decay and an effective batch of 512 structures on four GH200 GPUs, and the checkpoint with the lowest validation error is kept. The members are finally assembled into a single model in which the shared embeddings and first GNN layer are evaluated once, and whose output is the member mean. The variant branched at the heads follows the same procedure, with only the heads and readouts re-initialized and trained.

\section{Failure analysis of the IR and Raman spectra}
\label{sec:si_spectra_failures}

\subsection{Spectral intensities}
\label{sec:si_spectra_intensities}

A normal mode $k$ with mass-weighted, normalized eigenvector $\mathbf{e}_k$ displaces atom $i$ by $Q_k\, \mathbf{e}_{k,i} / \sqrt{m_i}$, so that $Q_k$ is in \AA\,amu$^{1/2}$. The derivatives $\boldsymbol{\mu}'_k = \partial\boldsymbol{\mu}/\partial Q_k$ and $\boldsymbol{\alpha}'_k = \partial\boldsymbol{\alpha}/\partial Q_k$ are taken by central differences at $Q_k = \pm 10^{-3}$\,\AA\,amu$^{1/2}$.
The IR intensity is
\begin{equation}
    I_k = \frac{N_\mathrm{A}}{12\,\varepsilon_0 c^2} \left| \boldsymbol{\mu}'_k \right|^2 ,
\end{equation}
in km\,mol$^{-1}$, and the Raman activity is
\begin{equation}
    S_k = 45\, a_k'^2 + 7\, \gamma_k'^2 ,
\end{equation}
in \AA$^4$\,amu$^{-1}$, with the isotropic and anisotropic invariants of $\boldsymbol{\alpha}'_k$
\begin{equation}
    a_k' = \tfrac{1}{3} \mathrm{tr}\, \boldsymbol{\alpha}'_k , \qquad
    \gamma_k'^2 = \tfrac{1}{2} \left[ (\alpha'_{xx} - \alpha'_{yy})^2 + (\alpha'_{yy} - \alpha'_{zz})^2 + (\alpha'_{zz} - \alpha'_{xx})^2 + 6 \left( \alpha'^2_{xy} + \alpha'^2_{yz} + \alpha'^2_{xz} \right) \right] ,
\end{equation}
the convention of the ORCA reference data. Spectra are sums of Lorentzians of these intensities centered at the reference frequencies.

\subsection{Exclusions}
\label{sec:si_spectra_exclusions}

Of the 7193 molecules of IR-R-7193, 9 are excluded from every score, for two reasons.
Four have an odd number of electrons (nitroxide radicals), as in Ref.~\onlinecite{gonnheimer2026macemdp}.
Five have unphysical or incorrect reference intensities: four with reference frequencies above 5000~cm$^{-1}$, IR intensities above $10^{4}$~km\,mol$^{-1}$ or Raman activities above $10^{5}$~\AA$^4$\,amu$^{-1}$, or flagged by inspection, and 2,6-difluorophenyl isothiocyanate, whose reference fails the check below.
For a further 47, the reference geometry does not correspond to the molecule it is labeled as. We build the bond graph of each reference geometry, hydrogens included, from covalent radii scaled by 1.2, and compare it by graph isomorphism, with atoms matched by element, with the structure listed in PubChem for the CAS number of the molecule.
The geometries that do not match are compared in the same way with the two-dimensional structure of the NIST Chemistry WebBook, and those that match neither are marked as mislabeled.
These geometries appear to have been inherited from three-dimensional structure files that contradict the two-dimensional structure of the same entry. The DFT spectra are consistent with the computed structures, so they are kept in the scores, where they carry many of the lowest-scoring spectra (10 of the 32 IR and 5 of the 29 Raman spectra of \texttt{PET-M} with $r_\mathrm{msc} < 0.95$), but they are not used as examples, and neither are six further geometries that fall apart into several bonded fragments.

\subsection{Reference check}

We recomputed reference intensities for seven molecules: four for which PET-M and MACE-MDP agree with each other better than with the reference, and three failures of PET-M, including azulene and tris(3-chlorophenyl)phosphine.
For each molecule we took the three modes with the largest PET--DFT difference and the strongest mode, and computed $\boldsymbol{\mu}$ and $\boldsymbol{\alpha}$ with $\omega$B97M-D3(BJ)/def2-TZVPPD, the level of theory of the reference, with polarizabilities from coupled-perturbed Kohn--Sham linear response rather than finite fields, and derivatives and intensities computed as for PET (Sec.~\ref{sec:si_spectra_intensities}).
Only 2,6-difluorophenyl isothiocyanate disagrees with its reference, by orders of magnitude for some Raman activities, and it is excluded.

\subsection{Local LLPR}\label{sec:si_llpr}

For each output block $b$ of PET-M ($\boldsymbol{\mu}$, and the $\lambda = 0$ and $\lambda = 2$ parts of $\boldsymbol{\alpha}$), the prediction is linear in the last-layer features $\mathbf{h}$, the sum of the atomic features $\mathbf{h}_i$.
With $\mathbf{A}_b = \sum_n \mathbf{h}_n \mathbf{h}_n^\top$ accumulated over the $1.49 \times 10^6$ training structures and $\varsigma = 10^{-6}\,\mathrm{tr}(\mathbf{A}_b)/F$, with $F$ the number of features, the uncertainty of atom $i$ is
\begin{equation}
    \sigma_i^2 = a_b^2\, \mathbf{h}_i^\top \left(\mathbf{A}_b + \varsigma \mathbf{I}\right)^{-1} \mathbf{h}_i,
\end{equation}
where the calibration factor $a_b = \sqrt{\pi/2}\, \langle |r| / \sigma \rangle$ is computed from the residuals $r$ of the 82\,853 validation structures.
Along each normal mode $Q_k$, the same expression applied to $\mathbf{g}_{ik} = [\mathbf{h}_i(Q_k = \delta) - \mathbf{h}_i(Q_k = -\delta)]/2\delta$, with $\delta = 10^{-3}$\,\AA\,amu$^{1/2}$ along the mass-weighted normal coordinate, as for the intensities, gives the uncertainty $\sigma_{ik}$ of the contribution of atom $i$ to the derivative.
Each molecule receives two scores: the largest $\sigma_i$ at the reference geometry, and the largest root mean square over modes of $\sigma_{ik}$, divided by the root mean square over modes of the predicted derivative $|\partial y/\partial Q_k|$.
For IR both use the $\boldsymbol{\mu}$ block.
For Raman the first uses the $\lambda = 2$ block, and the second combines the $\lambda = 0$ and $\lambda = 2$ blocks in quadrature.
Each score is converted to a percentile over the 7184 molecules, and the flag is the larger of the two.
\subsection{Orientation test}\label{sec:si_spectra_orientation}

Each molecule and its reference normal modes were rotated rigidly into 24 random orientations, the derivatives predicted in each orientation were rotated back, and the spectra were computed from each orientation and from the orientation-averaged derivatives, which are exactly equivariant.
We tested all spectra with $r_\mathrm{msc} < 0.98$ outside the mislabeled geometries (63 IR and 131 Raman) and 20 random spectra with $r_\mathrm{msc} > 0.99$ per channel as controls.
The median spread (maximum minus minimum) of $r_\mathrm{msc}$ over orientations is 0.018 (IR) and 0.025 (Raman) for the former, against 0.001 for the controls.
For the per-peak comparison, strong peaks are those with a DFT intensity above 10\% of the strongest one, in molecules with at least four of them (61 IR and 126 Raman spectra).
The most mispredicted peak is the one with the largest $|\ln(I_\mathrm{PET}/I_\mathrm{DFT})|$, and the orientation sensitivity of a peak is the standard deviation of its intensity over orientations divided by its mean.
The chance expectation is the sum over molecules of three divided by the number of strong peaks.
Because the sensitivity is divided by the mean predicted intensity, it is larger for under-predicted peaks, which also have the largest $|\ln(I_\mathrm{PET}/I_\mathrm{DFT})|$; the association persists, however, with the absolute standard deviation and on the spectra whose most mispredicted peak is over-predicted. With the absolute standard deviation, the association disappears for the control spectra, so it is not claimed for well-predicted spectra.
Since $\boldsymbol{\alpha}$ is a sum of atomic contributions, the equivariance error of atom $i$ along mode $k$ in Fig.~\ref{fig:spectra}d (right) is instead evaluated on the O(3) quadrature of Sec.~\ref{sec:si_quadrature}, as the weighted root mean square over its operations of the Frobenius norm of the deviation of $\partial\boldsymbol{\alpha}_i/\partial Q_k$ from its quadrature average.

\begin{table}
\caption{Agreement of the PET models and MACE-MDP with DFT on the 7184 scored molecules of IR-R-7193, at DFT normal modes. All values in \% except the mean absolute error (MAE) of the mode intensities, in km\,mol$^{-1}$ (IR) and \AA$^4$\,amu$^{-1}$ (Raman). $r_\mathrm{msc}$: mean over molecules (standard deviation in parentheses), over 0--4000\,cm$^{-1}$. MAE: over all normal modes, pooled. $\tilde{s}$: median ratio of total predicted to total reference intensity.}
\label{tab:si_spectra}
\begin{tabular}{llccc}
\toprule
& Model & $r_\mathrm{msc}$ & MAE & $\tilde{s}$ \\
\midrule
\multirow{4}{*}{IR} & XS & 99.4 (1.6) & 2.09 & 98.8 \\
 & S & 99.7 (1.6) & 1.49 & 99.0 \\
 & M & 99.7 (1.5) & 1.24 & 99.0 \\
 & MACE-MDP & 99.6 (1.8) & 1.68 & 100.7 \\
\midrule
\multirow{4}{*}{Raman} & XS & 99.5 (1.4) & 2.69 & 96.7 \\
 & S & 99.7 (1.2) & 2.31 & 95.1 \\
 & M & 99.7 (1.2) & 2.04 & 102.2 \\
 & MACE-MDP & 99.6 (1.3) & 2.49 & 100.2 \\
\bottomrule
\end{tabular}
\end{table}

\section{NMR chemical shielding tensors}
\label{sec:si_nmr}

All test metrics on ShiftML-molecules exclude 32 test structures, all conformers of six molecules with unusual nitrogen chemistry (N-nitroso, azo, cyanamide-like and imine groups, and a P--N--S ring) and reference $^{15}$N isotropic shieldings of up to 660~ppm, leaving 4105 test structures.
These were identified by $^{15}$N isotropic shielding errors of more than 80~ppm in at least three of six preliminary PET trainings, with all conformers of each molecule removed.
As the filter selects on the error of the target, it removes only nitrogen outliers, and changes the $^{1}$H and $^{13}$C metrics little.
The equivariance subset is the 990 of the remaining structures that fall in a random draw of 1000 test structures.
Four test structures, one molecule in four conformers, duplicate training structures; two of them fall in the equivariance subset, and all are kept for consistency with the external reference model.

\begin{figure}
    \centering
    \includegraphics[width=\linewidth]{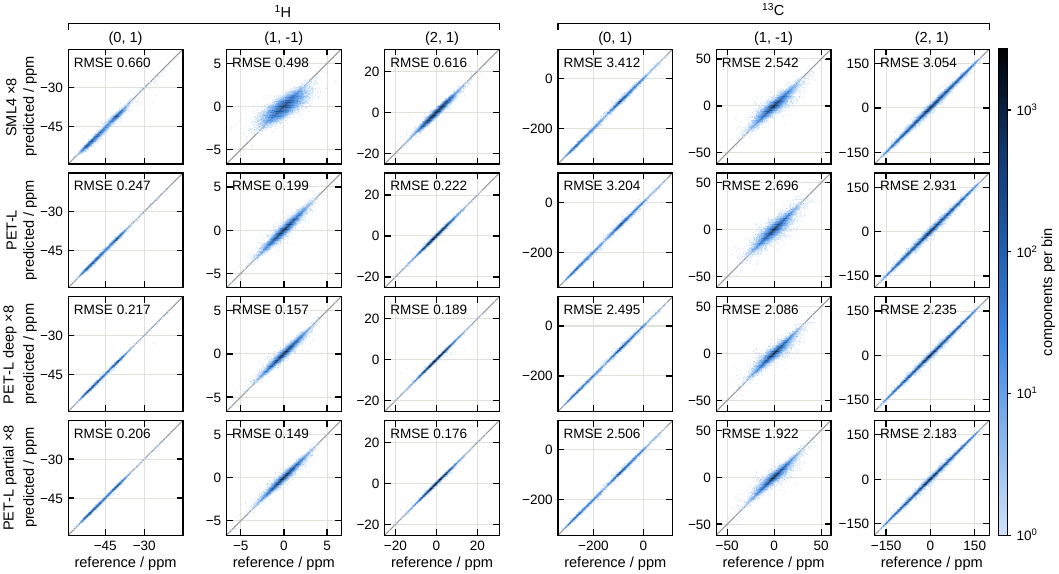}
    \caption{
    Parity of the predicted spherical components of the $^{1}$H and $^{13}$C shielding tensors against the reference, for each irrep $(\lambda, \sigma)$, in the original orientation of the 990 structures of the equivariance subset of the ShiftML-molecules test set. Rows: the 8-member ShiftML4 ensemble~\cite{kellner2026shiftml4}, and \texttt{PET-L} as a single model and as 8-member deep and partial ensembles. Each point is one spherical component of one atom; counts are binned in hexagons on a logarithmic color scale, and the RMSE over all points of each panel is given in ppm. Being from a single orientation, it differs slightly from the total RMSE of Table~\ref{tab:si_purification_hc}, which is averaged over the orientations of the grid of degree 7.
    }
    \label{fig:si_shiftml_parity_irreps}
\end{figure}

\begin{figure*}[t]
    \centering
    \includegraphics[width=\linewidth]{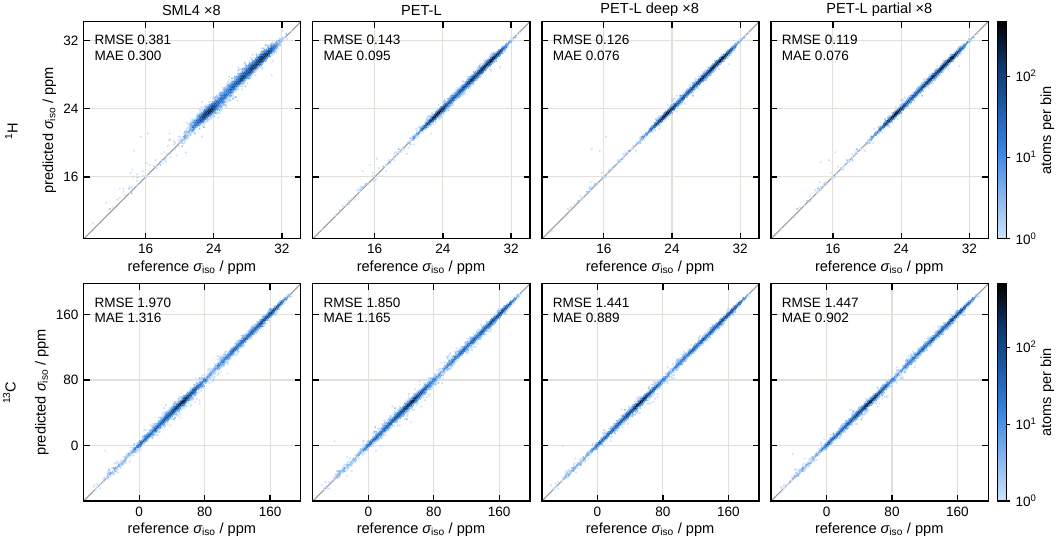}
    \caption{
    Parity of the predicted isotropic shielding $\sigma_{\text{iso}}$ of $^{1}$H (top) and $^{13}$C (bottom) nuclei against the reference, in the original orientation of the 990 structures of the equivariance subset of the ShiftML-molecules test set, for the 8-member ShiftML4 ensemble~\cite{kellner2026shiftml4} and for \texttt{PET-L} as a single model and as 8-member deep and partial ensembles. Counts are binned in hexagons on a logarithmic color scale, and the RMSE and MAE are given in ppm.
    }
    \label{fig:si_shiftml_parity_isotropic}
\end{figure*}

Figures~\ref{fig:si_shiftml_parity_irreps} and \ref{fig:si_shiftml_parity_isotropic} give the parity plots behind the comparison of the main text, for each irrep of the shielding tensor and for the isotropic shielding, for the ShiftML4 ensemble and for \texttt{PET-L} as a single model and as deep and partial ensembles.

\section{Orientation dependence of one-shot observables}
\label{sec:si_scfbench_o3}

The predicted auxiliary coefficients are not exactly equivariant, so the one-shot energy $E$ and dipole $\boldsymbol{\mu}$ derived from them need not be exactly invariant and covariant.
We test this for the three \texttt{PET-L} models i)--iii) of Fig.~\ref{fig:density_validation}a,b, on 50 molecules drawn at random from each of the 250-structure subsets of the in-distribution and out-of-distribution test sets (12--20 and 26--59 atoms).
Each molecule is transformed by 52 elements of $\mathrm{O}(3)$: for each of the 26 directions $\textbf{n}$ of the Lebedev rule of order 7, the rotation taking the $z$ axis to $\textbf{n}$ (the identity for $\textbf{n} = \hat{\textbf{z}}$) and its composition with inversion.
In each orientation the density is predicted in single precision, as in production, and $E$ and $\boldsymbol{\mu}$ are computed from the one-shot density matrix as in the main text, with $\boldsymbol{\mu}$ rotated back to the input frame.
The spreads are the standard deviation of $E$ and the root-mean-square deviation of the back-rotated $\boldsymbol{\mu}$ from its mean over the 52 orientations.
With a single in-plane angle per direction and no quadrature weights, these are spreads over a sample of orientations, not the weighted orbit variance of Sec.~\ref{sec:si_quadrature}, and may underestimate it.

\begin{figure*}[t]
    \centering
    \includegraphics[width=\textwidth]{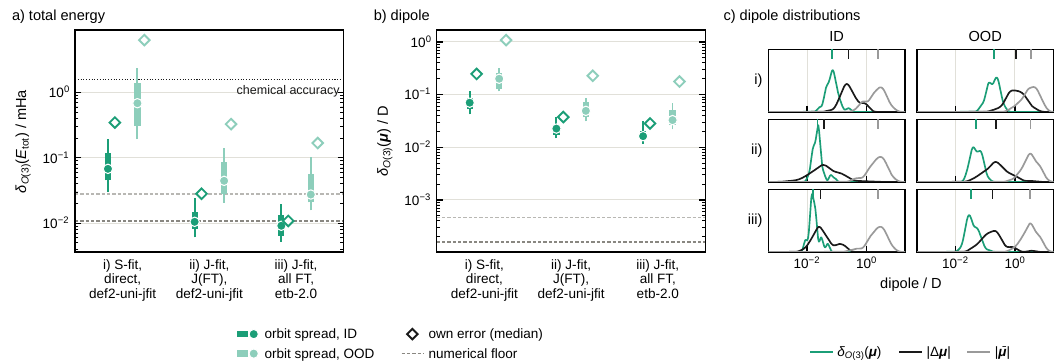}
    \caption{
    Spread over 52 orientations of the one-shot a) total energy and b) dipole of the SCFBench \texttt{PET-L} models i)--iii) of Fig.~\ref{fig:density_validation}, for 50 molecules from each of the in-distribution (dark) and out-of-distribution (light) test sets. Markers give the median, thick and thin bars the 25--75 and 10--90 percentiles over molecules. Open diamonds give the median one-shot error of the same model on the same molecules. Dashed lines give the numerical floor of each test set, from the exactly covariant \texttt{minao} guess density passed through the same one-shot path; the dotted line marks chemical accuracy. c) Distributions over the same molecules, for each model (rows) and test set (columns), of the dipole spread $\delta_{\mathrm{O}(3)}(\boldsymbol{\mu})$, the one-shot dipole error $|\Delta\boldsymbol{\mu}|$ and the magnitude $|\boldsymbol{\mu}|$ of the orientation-averaged predicted dipole, as kernel densities of their logarithm; ticks mark the medians.
    }
    \label{fig:si_scfbench_o3_variance}
\end{figure*}

Even an exactly equivariant density does not give exactly invariant one-shot observables, since the exchange-correlation potential is integrated on a molecular grid that does not rotate with the molecule.
We estimate this floor by passing the exactly covariant \texttt{minao} guess density of PySCF through the same one-shot path, on 10 in-distribution and 5 out-of-distribution molecules.
For the J-fitted models the energy spread is at this floor, with medians of 0.01~mHa in distribution and 0.03--0.04~mHa out of distribution, while the dipole spread of 0.02--0.05~D lies 70--140 times above it.
The S-fitted model trained directly on the coefficients varies more, with median spreads of up to 0.68~mHa and 0.20~D out of distribution.
For every model and test set, the median spread is below both chemical accuracy and the median one-shot error of the model on the same molecules (Fig.~\ref{fig:si_scfbench_o3_variance}a,b).
Molecule by molecule (Fig.~\ref{fig:si_scfbench_o3_variance}c), the dipole spread lies below the one-shot dipole error of the same molecule for 38 to 50 of the 50 molecules of each set, the exceptions mostly in distribution, where the J-fitted models are most accurate, and amounts to a median of 0.6 to 1.4\% of the predicted dipole for the J-fitted models and 3 to 5\% for the S-fitted one.

\section{Fine-tuning on external density datasets}
\label{sec:si_finetuning}

\subsection{Protocol}

All fine-tuning runs start from the J-fitted, J-loss \texttt{PET-L} models of the main text, in the \texttt{def2-universal-jfit} or \texttt{etb-2.0} basis, and leave every parameter trainable.
They use the J density loss on J-fitted targets, Adam with a 5\% linear warm-up followed by a cosine decay to zero, gradient clipping at norm 1.0, batches of eight structures and one GH200 GPU per run, at a peak learning rate of $10^{-4}$ for BFDb-SSI and $5\times10^{-5}$ for CAFES-QM9.
Each run uses a fixed budget of $2\times10^{4}$ optimizer steps, whatever the size $N$ of the training set, so that models at different $N$ differ only in the data they have seen, and the selected model is the one with the lowest validation loss; for CAFES-QM9 at $N = 10$ and 30 this checkpoint comes within the first few hundred steps. Each configuration is trained once, with a single random seed.
For BFDb-SSI, the atomic baseline and target scales are refitted on the training subset of each run, and the models trained from random weights use the same architecture and budget, at a peak learning rate of $2\times10^{-4}$; at $N = 1900$ both are additionally trained for $10^{5}$ steps.
For CAFES-QM9, as the auxiliary basis is the same and the readout layers are not re-initialized (and no models are trained from scratch), the baseline and scales of SCFBench are reused.

\subsection{BFDb-SSI}

\subsubsection{Dataset and splits}

The side-chain--side-chain interaction subset of the BioFragment Database, as prepared in Ref.~\onlinecite{briling2021metrics}, contains 2287 dimers of amino-acid side chains and 35 monomers, with 70 of the 322 test structures carrying a net charge (67 of $\pm 1$ and 3 of $\pm 2$), with $\omega$B97X-D/cc-pVQZ densities J-fitted on the cc-pVQZ-JKFIT auxiliary basis.
This basis reaches $\lambda = 5$ on C, N and O.
Because it differs from those of SCFBench, the readouts of the pretrained model cannot be reused: the backbone and the per-block heads for $\lambda \le 4$ are inherited, while the $\lambda = 5$ head and every readout start from random weights.
We train on nested subsets of between 30 and 1900 of the 2000 training structures of Ref.~\onlinecite{briling2021metrics}, keeping 100 for validation, and evaluate on its 322 test structures.
Ref.~\onlinecite{briling2021metrics} distributes the predictions of its SA-GPR models on the same test structures, which we evaluate with the same code as the PET predictions.

At $N = 1900$ the fine-tuned model has a Coulomb-metric error of 0.040~kcal/mol against 0.27~kcal/mol for the better SA-GPR model, an ESP error of 0.10 against 0.26~kcal/mol, and a dipole error of 0.019 against 0.060~a.u.
With $10^{5}$ steps at $N = 1900$ the fine-tuned model reaches a Coulomb-metric error of 0.016~kcal/mol and the from-scratch model 0.030~kcal/mol, with the same ordering for the other metrics.

\subsubsection{Electron-number correction}
\label{sec:si_electron_correction}

\begin{figure}[tb]
    \centering
    \includegraphics[width=\linewidth]{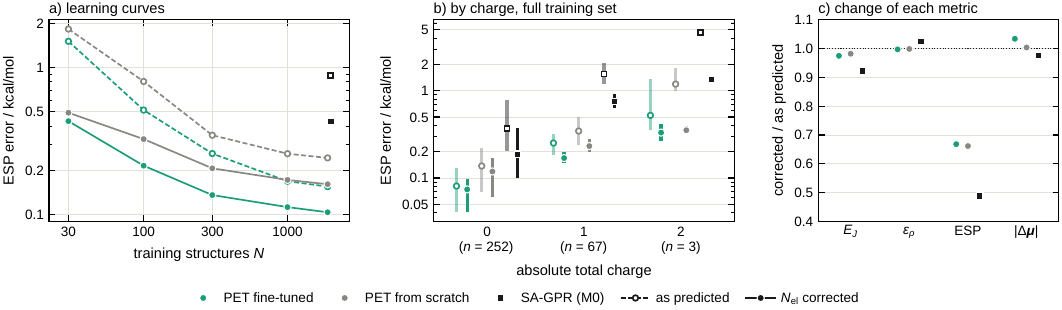}
    \caption{
    Effect of the a posteriori electron-number correction on the BFDb-SSI models ($2\times10^{4}$ steps). Hollow markers and dashed lines: predictions as they are; filled markers and solid lines: corrected.
    a) Learning curves of the ESP error. b) The same error at full training size, by absolute total charge of the test structure (marker: median; bar: 25--75\%, or the full range for the three doubly charged structures). c) Ratio of the corrected to the uncorrected test-set mean error for each metric.
    }
    \label{fig:bfdb_electron_correction}
\end{figure}

None of the models receives the total charge as an input.
The predicted densities therefore do not integrate to the correct number of electrons, and the error, of order $10^{-3}$ electrons for neutral and $10^{-2}$ for charged structures, is systematic: cations are predicted with too many electrons and anions with too few.
Following Ref.~\onlinecite{briling2021metrics}, we correct this \textit{a posteriori} by moving each predicted density to the closest one, in the Coulomb metric, that integrates to the exact electron count,
\begin{equation}
    \tilde{\textbf{c}}^{\text{aux}}_A \rightarrow \tilde{\textbf{c}}^{\text{aux}}_A + \kappa \, \textbf{J}_A^{-1} \textbf{q}_A ,
\end{equation}
where $\textbf{J}_A$ is the Coulomb metric $\hat{J}_A$ of Sec.~\ref{sec:si_losses}, $\textbf{q}_A$ collects the integrals of the auxiliary basis functions and the multiplier $\kappa$ is fixed by the electron count.
Figure~\ref{fig:bfdb_electron_correction} shows that this leaves the Coulomb-metric error, the density difference and the dipole essentially unchanged, and lowers the error of the electrostatic potential (ESP) by a third overall and by half for the charged structures, where a uniform offset of the potential is removed. The ESP is evaluated on a surface of very low density, about $2\times10^{-6}$~bohr$^{-3}$~\cite{briling2021metrics}, where it is dominated by the low multipoles of the density, which makes this metric particularly sensitive to the total charge.
All BFDb-SSI results are corrected in this way, as are the SA-GPR baselines and the fitted densities of the CAFES-QM9 results.

\subsubsection{Errors by total charge}

\begin{figure}[tb]
    \centering
    \includegraphics[width=\linewidth]{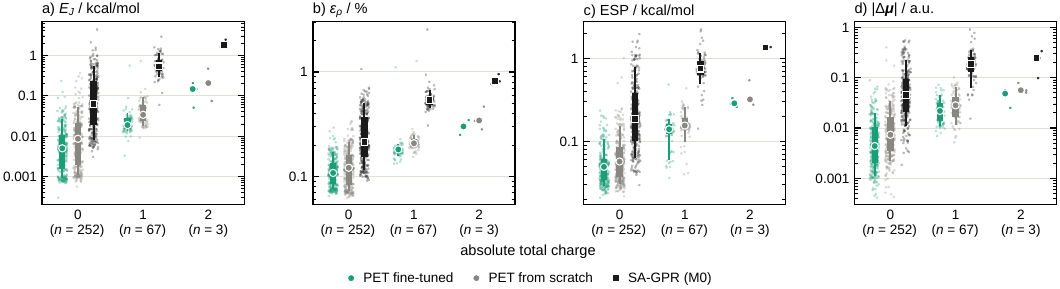}
    \caption{
    Test-set errors of the corrected BFDb-SSI models ($N = 1900$, $10^{5}$ steps) and of the corrected SA-GPR model, resolved by the absolute total charge of the test structure. Dots are individual structures; marker, thick and thin bars give the median and the 25--75\% and 10--90\% ranges (dots and median only for the three doubly charged structures).
    }
    \label{fig:bfdb_charge_errors}
\end{figure}

Figure~\ref{fig:bfdb_charge_errors} resolves the remaining error by the total charge of the test structure.
After correction, the median Coulomb-metric error of the fine-tuned model rises from 0.005~kcal/mol for neutral to 0.019 for singly and 0.15 for doubly charged structures, and the median ESP error from 0.05 to 0.14 and 0.29~kcal/mol, with SA-GPR degrading more steeply still.
The worst case of all three models is the same structure, a methylammonium monomer, one of only 25 monomers in the training set.

\subsection{CAFES-QM9: coupled-cluster densities from a DFT model}
\label{sec:si_cafes}

\begin{figure}[tb]
    \centering
    \includegraphics[width=\linewidth]{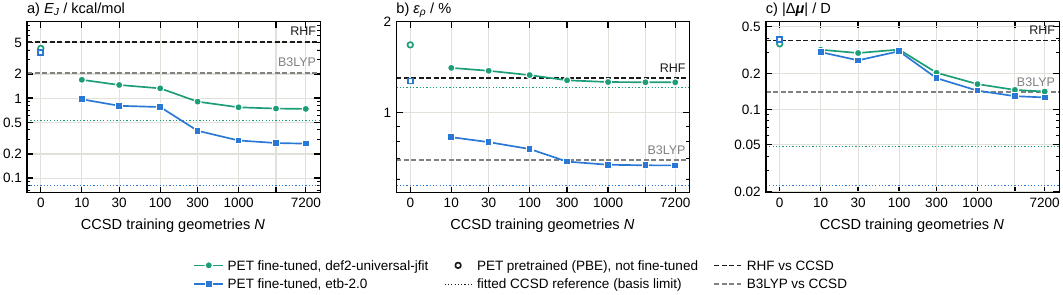}
    \caption{
    Fine-tuning the SCFBench \texttt{PET-L} models on CCSD/def2-SVP densities of QM9 molecules~\cite{li2026agentic}, J-fitted on the \texttt{def2-universal-jfit} (green) and \texttt{etb-2.0} (blue) auxiliary bases of the pretrained models, against the number $N$ of CCSD training geometries; hollow markers at $N = 0$ are the pretrained models as they are.
    Errors are against the exact CCSD density, averaged over the 2000 test geometries of the 100 molecules with 8 or 9 heavy atoms, which are held out entirely: a) the Coulomb-metric error $E_J$; b) the density difference $\epsilon_\rho$; c) the dipole moment error.
    Dashed lines: the RHF and B3LYP densities of the same geometries against CCSD. Dotted lines, in the basis color: the J-fitted CCSD reference itself, the limit of each auxiliary basis.
    Every fitted density carries the electron-number correction of Sec.~\ref{sec:si_electron_correction}.
    }
    \label{fig:cafes_finetuning}
\end{figure}

Here we ask whether a model pretrained on PBE densities can learn a correlated density from a few expensive calculations.
The CAFES-QM9 dataset~\cite{li2026agentic} provides all-electron, unrelaxed CCSD/def2-SVP one-particle density matrices for 500 QM9 molecules at 20 geometries each, sampled from \textit{ab initio} molecular dynamics, together with the Hartree--Fock and B3LYP densities of the same geometries.
As the orbital basis is that of SCFBench, the CCSD densities are J-fitted on the \texttt{def2-universal-jfit} and \texttt{etb-2.0} auxiliary bases with the same recipe as the reference data of the pretrained models, which are then fine-tuned on their own basis with nothing re-initialized.
We train on nested subsets of between 10 and 7200 geometries of the molecules with up to seven heavy atoms, stratified by parent molecule, select each model on 80 geometries of 40 further molecules of the same size, and test on all 2000 geometries of the 100 molecules with eight or nine heavy atoms, so that no test molecule is seen in training or validation and all are larger than those that are.
Because the density matrices are available, errors are measured against the exact CCSD density rather than its fit, which places the models, the other levels of theory and the auxiliary bases themselves on one scale (Fig.~\ref{fig:cafes_finetuning}).

In the Coulomb metric, the pretrained PBE model has an error of 4.2~kcal/mol against CCSD, about as large as that of Hartree--Fock (5.1~kcal/mol) and twice that of B3LYP (2.1~kcal/mol). This error combines the difference between the PBE and CCSD densities, which we have not computed for these geometries, with the error of the model on its own PBE target.
Ten CCSD geometries, one per molecule, lower it below B3LYP, to 1.7 and 1.0~kcal/mol in the \texttt{def2-universal-jfit} and \texttt{etb-2.0} bases respectively, and 300 geometries bring the \texttt{etb-2.0} model to 0.39~kcal/mol, a factor of five below B3LYP and within a factor of five of the limit of its basis.
The real-space density is limited by the auxiliary basis instead.
The RI floor of \texttt{def2-universal-jfit} is $\epsilon_\rho = 1.21\%$, close to the Hartree--Fock value of 1.30\%, and every model in this basis sits just above it, while in the \texttt{etb-2.0} basis, with a floor of 0.57\%, the fine-tuned model reaches the B3LYP value (0.69\%) with 300 geometries and 0.67\% with 1000 or more.
The dipole moment error approaches the B3LYP level of 0.14~D with 1000 geometries, falls slightly below it in the \texttt{etb-2.0} basis with 3000 or more (0.13~D), and reaches it in \texttt{def2-universal-jfit} only with all 7200, although both floors lie far below, so that here the limit is the model.
The Coulomb metric, which the model is trained on and which governs the potential the density generates, is thus learned to well beyond DFT accuracy from a handful of CCSD calculations.

\section{Inference cost and memory}
\label{sec:si_timing}

All models in Fig.~\ref{fig:pareto} are timed on one NVIDIA GH200 GPU. For each model, the largest batch that fits in memory is found on the test molecules of its dataset, and the inference time per atom is measured at that batch size.
The timings use the full test set of each dataset (the filtered one for ShiftML-molecules, and the in-distribution one for SCFBench), cycled to fill every batch, whereas the accuracies are those of the main text: the full test set for SPICE-$\alpha$, the 990-structure subset for ShiftML-molecules, and the 250 in-distribution structures of Fig.~\ref{fig:density_validation} for SCFBench. Batch sizes are counted in structures. For each model, a descending search over powers of two followed by three bisection steps finds the largest batch that runs without running out of memory and without allocator retries, which stall individual batches close to the memory limit. Where the test set is too small for this search to reach the limit (SCFBench \texttt{PET-XS} and \texttt{PET-S}), the capacity is measured separately on batches built by repeating test structures, and the timing is taken at the largest batch the test set allows, where the time per atom has already converged. Each timing run has 10 warm-up and 8 timed batches (3 and 4 for the 8-member deep ensembles), and the reported time is the mean over three independent runs. Only the forward pass is timed, on a batch already on the GPU, including the heads and readouts; neighbor lists are computed outside the timed region for every model. All models run in single precision, except MACE-MDP, in double precision as released, and a double-precision \texttt{PET-M} comparison point; no mixed precision or compilation is used. The software is PyTorch 2.9.1 with CUDA 12.9; MACE-MDP from its repository with mace-torch 0.3.15 and e3nn 0.4.4, without cuEquivariance kernels; the released ShiftML4 models, re-exported without changing their weights; and NequIP-L from the SCFBench release with nequip 0.19.1 and e3nn 0.6.0.

The members of a deep ensemble are evaluated one after another on the same GPU, and their predictions averaged. As the intermediate tensors of each member are freed before the next one runs, a deep ensemble holds almost as many atoms per GPU as its single model: 66\,856 against 71\,553 for \texttt{PET-L}, and the same number for ShiftML4. Every atom is evaluated once per member, so a deep ensemble of 8 costs up to 8 times its single model; the measured ratios, 6.8 for \texttt{PET-L} and 5.4 for ShiftML4, are lower because the per-call work outside the members' forward passes is done once per batch, and is a larger share for the smaller ShiftML4 members. The members are independent until the final average, so they can equally run in parallel on separate GPUs at the time of a single model. In the partial ensemble the eight branches run together after the shared first GNN layer, which makes it faster than the deep ensemble (0.38 against 0.62~ms per atom) but holds the activations of all branches at once, reducing its capacity to about a third (22\,612 atoms). In Fig.~\ref{fig:equivariance_overview}d the cost of a ShiftML4 ensemble of $N$ members is taken as $N$ times that of one member, as shipped.

\end{document}